\pdfoutput=1

\documentclass[11pt]{article}
\usepackage{jheppub}
\usepackage{amsmath,amsfonts,latexsym,amssymb,slashed}
\usepackage{graphicx}
\usepackage{bbm}
\newcommand{\ft}[2]{{\frac{#1}{#2}}}
\usepackage{float}

\usepackage{caption}
\usepackage{subcaption}
\usepackage{tikz}
\usepackage{pgfplots}
\usetikzlibrary{matrix,arrows,decorations.pathmorphing}
\usetikzlibrary{patterns}
\usetikzlibrary{shapes.misc}
\usetikzlibrary{trees}
\usetikzlibrary{decorations.pathmorphing}
\usetikzlibrary{shapes.geometric}
\usetikzlibrary{positioning}
\usetikzlibrary{decorations.markings}
\usetikzlibrary{positioning,arrows}

\usetikzlibrary{decorations.pathreplacing}
\usetikzlibrary{decorations.pathmorphing}
\usetikzlibrary{shapes}
\usetikzlibrary{plotmarks}
\usetikzlibrary{decorations.markings}
\tikzset{
particle/.style={thin,draw=black, postaction={decorate},
decoration={markings,mark=at position .5 with {\arrow[black, line width=0.5mm]{stealth}}}},
gluon/.style={decorate, draw=black, decoration={coil,amplitude=4pt, segment length=5pt}},
photon/.style={decorate, decoration={snake}}
}
\newcommand{\cL}{\mathcal{L}}
\newcommand{\cM}{\mathcal{M}}
\newcommand{\cN}{\mathcal{N}}

\newcommand{\be}{\begin{equation}}
\newcommand{\ee}{\end{equation}}
\newcommand{\ba}{\begin{eqnarray}}
\newcommand{\ea}{\end{eqnarray}}

\renewcommand{\d}{\textrm{d}}

\renewcommand{\a}{\alpha}
\renewcommand{\b}{\beta}

\def\rmi{{\rm i}}

\newcommand{\rf}[1]{(\ref{#1})}
\newcommand{\bea}{\begin{eqnarray}}
\newcommand{\eea}{\end{eqnarray}}

\def\bfzero{\relax{\rm I\kern-.18em 0}}
\def\bfone{\relax{\rm 1\kern-.35em 1}}
\def\twomat#1#2#3#4{\left(\begin{array}{cc}
\end{array}
\right)}

\def\a{\alpha}
\def\b{\beta}

\def\eps{\epsilon}
\def\g{\gamma}

\def\d{\delta}

\def\l{\lambda}

\def\g{\gamma}

\def\>{\rangle} %right angle
\def\<{\langle} %left angle
\def\l{\lambda}
\newcommand{\dslash}{\slashed{\partial }}
\def\pa{\partial}

\def\rmi{{\rm i}}
\def\rmd{{\rm d}}
\def\rme{{\rm e}}

\newcommand{ \chirlam}[1]{\langle #1\rangle}
\newcommand{ \achirlam}[1]{[#1]}

\newcommand{\SU}{\mathop{\rm SU}}
\newcommand{\SO}{\mathop{\rm SO}}
\newcommand{\U}{\mathop{\rm {}U}}

\newcommand{\cc}{{\rm c.c.}}
\newcommand{\hc}{{\rm h.c.}}

\newcommand{\ada}{A}
\newcommand{\adb}{B}
\newcommand{\adc}{C}
\newcommand{\add}{D}
\newcommand{\ade}{E}
\title  {Absence of $\U(1)$ Anomalous Superamplitudes in $\cN\geq 5$ Supergravities }
\author[a,b]{Daniel Z. Freedman,}
\author[a]{Renata Kallosh,}
\author[a]{Divyanshu Murli,}
\author[c]{Antoine Van Proeyen,}
\author[a]{and Yusuke Yamada}
\affiliation[a]{Stanford Institute for Theoretical Physics and Department of Physics, \\ Stanford University, Stanford, CA 94305, USA}
\affiliation[b]{Center for Theoretical Physics and Department of Mathematics, \\ Massachusetts Institute of Technology, Cambridge, MA 02139, USA}
\affiliation[c]{KU Leuven, Institute for Theoretical Physics, \\ Celestijnenlaan 200D, B-3001, Leuven, Belgium}
\emailAdd{dzf@math.mit.edu}
\emailAdd{kallosh@stanford.edu}
\emailAdd{divyansh@stanford.edu}
\emailAdd{antoine.vanproeyen@fys.kuleuven.be}
\emailAdd{yusukeyy@stanford.edu}

\notoc

\abstract{   We list all
potential candidates for $\U(1)$ anomalous non-local 1-loop 4-point amplitudes and higher loop UV divergences in $\cN\geq 5$ supergravities. The relevant  chiral superinvariants are  constructed from linearized chiral superfields and define the corresponding superamplitudes.
The  anomalous amplitudes, of the kind present in $\cN=4$, are shown to be absent in $\cN\geq 5$.  In $\cN=6$ supergravity  the result is deduced from the double-copy $(\cN=4)_{YM} \times (\cN=2)_{YM}$ model, whereas in $\cN=5,8$ the result on absence of anomalous amplitudes is derived in supergravities as well as in the $(\cN=4)_{YM} \times (\cN-4)_{YM}$ double-copy models.}

\keywords{Extended Supersymmetry, Scattering Amplitudes, Superspaces}
\arxivnumber{1703.03879}
\begin{document}

\maketitle

\newpage

 \tableofcontents{}

\newpage

%%%%%%%%%%%%%%%%%%%%%%%%%%%%%%%%%%%%%%%%%%%%%%%%%%%%%%%%%%%%%%%%%%%%%%
%%%%%%%%%%%%%%%%%%%%%%%%%%%%%%%%%%%%%%%%%%%%%%%%%%%%%%%%%%%%%%%%%%%%%%
%%%%%%%%%%%%%%%%%%%%%%%%%%%%%%%%%%%%%%%%%%%%%%%%%%%%%%%%%%%%%%%%%%%%%%
%%%%%%%%%%%%%%%%%%%%%%%%%%%%%%%%%%%%%%%%%%%%%%%%%%%%%%%%%%%%%%%%%%%%%%
%%%%%%%%%%%%%%%%%%%%%%%%%%%%%%%%%%%%%%%%%%%%%%%%%%%%%%%%%%%%%%%%%%%%%%
\section{Introduction}
Anomalies in extended supergravities and their relation to UV divergences have been studied since the  discovery of these theories~\cite{Christensen:1978gi,Christensen:1978md,Kallosh:1979au,Christensen:1979qj,Kallosh:1979pd}.  It has been known for a long time~\cite{Marcus:1985yy}  that chiral anomalies associated with topological Atiyah-Singer index theorem are absent for $\cN\geq 5$. Conformal anomalies associated with the Euler number, $\sum_s a_s$ and the Weyl anomaly, $\sum_s c_s$, were studied in~\cite{Nicolai:1980td,Duff:1982yw,Fradkin:1982bd,Fradkin:1981jc,Fradkin:1985am,Duff:1993wm}. A recent surprise  was a  discovery in~\cite{Meissner:2016onk} that in the harmonic gauge for gauge fields  the Weyl conformal anomalies, $\sum_s c_s$, vanish for $\cN\geq 5$.
 A possible supersymmetric explanation of the absence of chiral and Weyl anomalies in $\cN\geq 5$ was proposed in~\cite{Kallosh:2016xnm}.
 In $\cN=4$ supergravity the chiral anomaly is present~\cite{Marcus:1985yy}. More recently it was discovered in \cite{Carrasco:2013ypa} that there are anomalous 4-point amplitudes in $\cN=4$ supergravity that are supersymmetric but violate helicity/chirality conservation.  Supergravity amplitudes are constructed as the product of gauge theory amplitudes  using the double-copy method~\cite{Bern:2008qj,Bern:2010ue,Elvang:2015rqa}.  The anomalous amplitudes first arise as (UV finite and nonlocal) contributions at the 1-loop level.  From their associated superamplitudes,  two independent candidate local 4-loop counterterms were constructed in addition to the conventional  counterterm that contains the MHV 4-graviton amplitude and its SUSY partners.

The $\cN =4$ theory is finite at 3-loop order, \cite{Bern:2012cd}, but the 4-loop calculations in~\cite{Bern:2013uka} revealed that the $\cN=4$ theory contains the predicted UV divergent structures.
The superamplitude in double copy form is\footnote{Here and below we use the normalization of the external states in the 4-point amplitude in agreement with~\cite{Kallosh:1980fi}, where all dependence on $\kappa$, for example in the Einstein action, is given by ${\frac12 \kappa^2} R$.}
 \be
{\cal M}_{\cN=4, \rm \,  div}^{\rm 4-loop}=  \frac{C_4}{\epsilon}  stA^{\rm tree}_{\cN=4 \rm YM} ( {\cal O}^{- -+ +} + 60\,   {\cal O}^{+ ++ +} - 3 \,{\cal O}^{- + + +} )
\label{zvi2}\ee
where $s=s_{12}, \, t=s_{23}, \, u=s_{13}$ and
\be
C_4= \frac{1}{(4\pi)^8} \Big ( \frac{\kappa}{2}\Big)^{6} \frac1{288} ( 264 \, \zeta_3-1)
\ee
The three terms ${\cal O}^{--++}$, etc. in the second factor indicate gluon helicity amplitudes in the $\cN=0$ Yang-Mills factor of the double copy.  The first term preserves helicity , while ${\cal O}^{++++}$ and ${\cal O}^{-+++}$ violate helicity conservation.  This is permitted since the $\cN=0$  gauge theory is non-supersymmetric.  Yet the product superamplitude in (\ref{zvi2}) is supersymmetric because  $A^{\rm tree}_{\cN=4 \rm YM}$ contains the well known factor $\delta^{(8)}(Q_{i \alpha})$, with $i=1,\ldots 4$ and $\alpha $ the chiral spinor index.  The superamplitude includes  the helicity non-conserving processes   $\<h^{++}h^{--}h^{++}\phi\>,~\<h^{++}h^{++}\phi\phi\>$ and their conjugates (plus their SUSY partners) in addition to the conventional process $\<h^{--}h^{--}h^{++}h^{++}\>.$

Four-loop computations in $\cN=5$ supergravity were reported in \cite{Bern:2014sna}.  The divergent diagrams can be expressed as the sum of the $\cN=4$ superamplitude of \rf{zvi2} plus an additional term proportional to it with equal and opposite coefficient. Thus
\be
   {\cal M}_{\cN=5, \rm \,  div}^{\rm 4-loop}=  {\cal M}_{\cN=4, \rm \,  div}^{\rm 4-loop} + \Delta {\cal M}_{ \rm \,  div}^{\rm 4-loop} = 0,
\ee
so all three terms in \rf{zvi2} cancel in $\cN=5$.
The additional contribution in $\cN=5$ supergravity,  $\Delta {\cal M}_{ \rm \,  div}^{\rm 4-loop}$, is due to the presence of the fermion in the double-copy $(\cN=4)_{YM} \times (\cN=1)_{YM}$ computations versus $(\cN=4)_{YM} \times (\cN=0)_{YM}$. The cancellation of helicity/chirality violating  terms in the UV divergences of $\cN=5$ might be due to absence of anomalous amplitudes in $\cN=1$ SYM theory, however, the reason why the 4-graviton helicity preserving amplitude is also finite remains a puzzle.

The situation described above suggests that it is important to study the issue of helicity/chirality violating amplitudes in $\cN\geq 5$ supergravity. Our goal  is to find and analyze anomalous amplitudes in $\cN\geq 5$  analogous to those found in $\cN=4$ supergravity in \cite{Carrasco:2013ypa}. This  requires  knowledge of the linearized chiral superfields in these theories. They have been already used in \cite{Kallosh:2016xnm} for the supersymmetric analysis of the chiral and conformal anomalies, but here we  systematically present the properties of these superfields, prove that they are chiral and  that there are no other chiral superfields besides those that we identify.

The main conclusions of our work are that there are chiral superfields and chirality/ helicity violating supersymmetric invariants for all $\cN$-extended supergravities that we classify in a simple systematic pattern.\footnote{An analogous classification for SUSY gauge theories is also given in appendix~\ref{ss:N4321}.} This means that there are candidate chirality violating amplitudes.  However,  in
$\cN=5$, 6, 8 supergravities,  closer study shows that these  potentially anomalous superamplitudes have vanishing coefficients.   This provides further evidence that the chiral anomaly present for $\cN=4$ and absent for
$\cN=5$, 6, 8 plays a significant role.

\section{Review of anomalous superamplitudes in \texorpdfstring{$\cN=4$}{N=4} supergravity}
\label{review}
We present further information on the results of  \cite{Carrasco:2013ypa} for $\cN=4$ supergravity because it sets the pattern for the extension to $\cN \geq 5$ in later sections of this paper.  The on-shell fields of the theory are (with Weyl spinor indices and $i=1,2,3,4$)
\begin{equation}
  \left\{{\cal N}=4\right\}= \left\{C_{\a\b\g\delta},\ \psi_{\a\b\g i},\  M_{\a\b ij},\ \chi_{\a}^{ i},\ \phi\right\}+ \cc\,.
\label{N4fields}
\end{equation}
These fields are components of two linearized on-shell chiral superfields  $\bar{C}_{\dot\a \dot\b \dot\g \dot\d}(y, \theta)$ and  $W(y, \theta) = \phi(y, \theta)$.  Here we present a schematic simplified version of these superfields, their complete form can be found in appendix~\ref{app:linchirSG}.
\begin{align}
\bar{C}_{\dot\a\dot\b\dot\g\dot\d}(y, \theta)= \bar C_{\dot\a\dot\b\dot\g\dot\d}(y) +\theta^{\a}_i  \partial_{\a (\dot{\a} } \bar{\psi}^i_{\dot \b \dot \g \dot \d)}+
 \dots + \ft1{4!}\theta^{\a}_i \theta^{\b}_j \theta^{\g}_k \theta^{\d}_{\ell } \partial_{\a \dot\a} \partial_{\b \dot\b}  \partial_{\g \dot \g} \partial_{\d \dot \d}\bar \phi \varepsilon^{ijk\ell }
 \,,
  \label{chiralbarCN4short}
\end{align}
\bea
W(y, \theta) &=& \phi(y)+ \theta^{\a}_i  \chi_\alpha ^i + \dots +\ft1{4!}\theta^{\a}_i \theta^{\b}_j \theta^{\g}_k \theta^{\d}_{\ell } C_{\a\b\g\delta} \varepsilon^{ijk\ell }\,.
\label{W}\eea
The conjugate fields to those in (\ref{N4fields}) are components of anti-chiral superfields.

We assign a $\U(1)_h$ quantum number to each superfield, which is equal to the helicity of its lowest component and under which $\theta^\a_i\to \rme^{\rmi b/2} \theta^\a_i$ (so that $\theta ^\alpha _i$ carries helicity $1/2$).  This $\U(1)_h$ acts on the physical states of the theory.  Anomalous amplitudes thus violate helicity conservation.\footnote{Later we will define a $\U(1)_c$ quantum number related to the $\U(1)_R$ subgroup of the duality group.}

The standard linearized 4-point supersymmetric $L$-loop invariants, which preserve helicity, are of the following form  \cite{Kallosh:1980fi,Howe:1980th}
\be
\cM_{\rm div} = \kappa^{2(L-1)} \int \rmd^4 x\, \rmd^{8} \theta \, \rmd^{8} \bar \theta\, \partial^{2(L-3)} W^2 \, \bar W^2\,,
\label{I}\ee
where $ \partial^{2(L-3)}$ is a symbolic expression representing  various contractions of spacetime derivatives acting on individual superfields.  We are interested in supersymmetric invariants for possible anomalous 1-loop 4-particle processes and thus look for $\rmd^8\theta$ integrals that are quartic in the superfields. There are two invariants. One is non-local, the other one is local.
In momentum space with $P=\sum_I p_I$,
\bea\label{M1}
\cM^{(1)}&=&\frac{\delta^4(P)}{stu}   \int \rmd^8  \theta \,  \bar C_{\dot \alpha \dot \beta\dot \gamma\dot \delta} (p_1,  \theta)  W (p_2, \theta)
 p_3 ^{\alpha \dot \alpha} p_3 ^{\beta \dot \beta}  W (p_3,  \theta)  p_{4 \alpha}{}^{\dot \gamma} p_{4\beta }{} ^{\dot \delta}  W (p_4,  \theta)   \,,\\
\cM^{(2)} &=&\delta^4(P) \int \rmd^8\theta\, W(p_1,\theta)W(p_2,\theta)W(p_3,\theta)W(p_4,\theta) \,. \label{M2}
\eea
Since $\rmd^8\theta$ carries $\U(1)_h$-charge $-4$, and $\bar C$ charge 2,  we see that $\cM^{(1)}$ and $\cM^{(2)}$  carry net helicity $-2$ and $-4$, respectively.  There is a $\theta^8$ term in the integrand of $\cM^{(1)}$ that corresponds to the process\footnote{We associate $\bar C_{\dot \alpha \dot \beta \dot \gamma \dot \delta }$ to $h^{++}$ and $C_{\alpha \beta \gamma \delta }$ to $h^{--}$.}
$\<h^{+ +} h^{- -}h^{- -} \phi\> $ that violates helicity conservation, while $\cM^{(2)}$ contains the process
$\<h^{--} h^{--}\phi\phi\>$. Of course, there are other helicity violating processes  related by SUSY to those above.  In both cases the anomalous effective action is given by $\cM +\overline\cM$.

 The corresponding super-amplitudes were computed in \cite{Carrasco:2013ypa} by the 1-loop double-copy method for $(\cN=4)_{YM} \times (\cN=0)_{YM}$ model  with the results (see the notation in appendix~\ref{app:conventions}) and in particular~(\ref{del2NQ})
  \bea
{\cal M}_{1-\rm loop }^{(1)}(1,2,3,4)=\frac{\rmi}{(4\pi)^2}\frac{1}{\chirlam{31}\chirlam{14}}\frac{\chirlam{32}\chirlam{24} \achirlam{21}}{\chirlam{21}}
\delta^8(Q) \delta^4 (P)\,,
\label{4ptnonlocal_app}
\eea
for the case of \rf{M1}, and for the case of \rf{M2}
\bea
{\cal M}_{1-\rm loop }^{(2)}(1,2,3,4) = -\frac{\rmi}{(4\pi)^2}\delta^8(Q)  \delta^4 (P) \ .
\label{III}\eea
Both expressions above have dimension zero and correspond to $\kappa$-independent 1-loop amplitudes.

We now observe that if we multiply by $\kappa^6 stu$ we arrive at local expressions which are supersymmetric and still dimensionless. They are therefore candidates for 4-loop divergences, in addition to the helicity preserving invariant in \rf{I} for $L=4$.  The helicity violating candidate for 4-loop divergences  can be obtained from the chiral invariants
\bea \label{M1div}
( \cM + \overline \cM)^{(1)}_{\rm div} &=& \frac{\kappa^6}{\epsilon} \int \rmd^4x\, \rmd^8  \theta\,[   \bar C_{\dot \alpha \dot \beta\dot \gamma\dot \delta}   W
 \partial^{\alpha \dot \alpha} \partial^{\beta \dot \beta}  W \partial^{\dot \gamma}{}_{ \alpha} \partial^{\dot \delta}{}_{ \beta}W]  +\hc \,,\\
( \cM + \overline \cM)^{(2)}_{\rm div} &=&\frac{\kappa^6}{\epsilon} \int \rmd^4x\, \rmd^8  \theta\, \pa^6 W^4 +\hc\,, \label{M2div}
\eea
where $\pa^6$ indicates the distribution of spacetime derivatives that produces $stu$ in Fourier space.
It is quite remarkable that the 4-loop calculations of  \cite{Bern:2013uka}  correspond exactly to these structures with coefficient given in \rf{zvi2}!

\

{\it Chiral $\U(1)_c$ symmetry and its anomaly}

\

\noindent In $\cN=4$ supergravity we  define the anomalous chiral $\U(1)_c$ symmetry as follows: it acts on chiral and anti-chiral linearized superfields as well as on the field components  and on $\theta$'s as follows.
\begin{eqnarray}
W(y, \theta_i) &\rightarrow& \rme^{ 2\rmi a}  W(y, \theta_i)\, ,  \qquad  \hskip 1.8 cm  \overline W(\bar y, \bar \theta^i) \rightarrow  \rme^{ -2\rmi a} \overline W(\bar y, \bar \theta^i), \cr
\theta_i &\rightarrow&  \rme^{ \frac{\rmi}{2} a} \theta_i \, ,  \qquad    \hskip 3.4 cm  \bar \theta^i \rightarrow  \rme^{ -\frac{\rmi}{2} a} \bar \theta^i.
\label{chiral}\end{eqnarray}
 Here $y^{\alpha \dot \alpha } = x^{\alpha \dot \alpha }+\ft12\bar \theta^{\dot \alpha i} \theta ^\alpha _i$, see (\ref{chiralDbasis}). The symmetry acting on the superfield, for example  $W(y, \theta_i) \rightarrow  \rme^{ 2\rmi a} W(y, \theta_i)$, means the following. We require that
 $\phi(y)  \rightarrow  \rme^{ 2\rmi a} \phi(y)$, $\theta_i \rightarrow  \rme^{ \frac{\rmi}{2} a} \theta_i$, $\chi_\alpha^i (y)  \rightarrow  \rme^{ \frac{ 3\rmi}{2} a}  \chi_\alpha^i (y)$, etc. so that every term in  \rf{W} transforms as the first one when both $\theta$ and higher components from spin 1/2 to spin 2,  being functions of $y$, transform accordingly. For example, the last term $\ft1{4!}\theta^{\a}_i \theta^{\b}_j \theta^{\g}_k \theta^{\d}_{\ell } C_{\a\b\g\delta} \varepsilon^{ijk\ell }$ satisfies this rule since $C_{\a\b\g\delta} $ is neutral under $\U(1)_c$ symmetry and the required factor $\rme^{ 2\rmi a}$ originates from 4 powers of $\theta$ in front of $C_{\a\b\g\delta} $ in the superfield.

The particular form of the $\U(1)$ symmetry in \rf{chiral} corresponds to the action of a $\U(1)_c$ normal subgroup of the isotropy group $H$ in the $\frac GH= \frac{\SU(1,1)\times \SO(6)}{\U(4)}$ \, coset space of $\cN=4$ pure supergravity. It will be generalized for all $\cN\geq 5$ models below.

The   4-graviton L-loop UV divergence in \rf{I} is invariant under the chiral transformations~\rf{chiral}. The other two UV divergences, in \rf{M1div} and in \rf{M2div},  are not invariant under
\rf{chiral}, for example
\be
\delta _{\U(1)_c} ( \cM + \overline \cM)^{(2)}_{\rm div} =  4 a \rmi( \cM - \overline \cM)^{(2)}_{\rm div}\,.
\label{superspaceUVch}
\ee

\

{\it Helicity conservation and its $\U(1)_h$ anomaly}

\

\noindent The chiral $\U(1)_c$ anomaly described above has a geometric origin associated with the isometry group $G$ and in particular with a $\U(1)$ normal subgroup of its isotropy subgroup $H$. On the other hand the 1-loop anomalous 4-point amplitudes  as well as the `extra' 4-loop UV divergences also break helicity conservation. The relation between these two types of anomalies is simple. We define here another $\U(1)_h$ transformations of the superfields and its components such that the corresponding chiral weight is a helicity of its first component, namely in $\cN=4$ case we define $\U(1)_h$ as follows
 \bea
W(y, \theta_i) &\rightarrow&  W(y, \theta_i)\, ,  \qquad  \hskip 1.8 cm  \overline W(\bar y, \bar \theta^i) \rightarrow  \overline W(\bar y, \bar \theta^i), \cr
\bar C_{\dot \alpha \dot \beta \dot \gamma\dot \delta}(y, \theta_i) &\rightarrow &  \rme^{ 2\rmi b}\bar C_{\dot \alpha \dot \beta\dot \gamma\dot \delta}(y, \theta_i) \, ,  \qquad  C_{ \alpha  \beta  \gamma \delta} (\bar y, \bar \theta^i) \rightarrow   \rme^{- 2\rmi b} C_{ \alpha  \beta  \gamma \delta} (\bar y, \bar \theta^i),\cr
\theta_i &\rightarrow&  \rme^{\ft12\rmi b} \theta_i \, ,  \qquad    \hskip 3.4 cm  \bar \theta^i \rightarrow  \rme^{ -\ft12\rmi b} \bar \theta^i\,.
\label{helicity}\eea
The difference with \rf{chiral} reflects the fact that the scalars have  vanishing helicity, but transform under the $G$ isometry group and its $\U(1)_c$ subgroup,
whereas the graviton is neutral in $G$ and its $\U(1)_c$ subgroup, but has a non-vanishing helicity $\pm 2$. Vectors participate in both $\U(1)$'s as well as spinors, although spinors helicity and chirality are different.

Thus, under chirality $\U(1)_c$ in \rf{chiral} and under helicity $\U(1)_h$ in \rf{helicity} the supersymmetry invariants transform as follows. This first one, $\cM_{\rm div}$,  is invariant under both $\U(1)$'s.
The second and the third one are both not invariant and we find that
\be
\delta _{\U(1)_c} ( \cM + \overline \cM)^{(1),(2)}_{\rm div} =  - \delta _{\U(1)_h}(\cM + \overline \cM)^{(1),(2)}_{\rm div}. \label{IchHelI}\ee

\section{Pure Extended  Supergravities}
\label{ss:pureExtSG}
In extended $\cN \geq 4$ pure (no matter) supergravities  scalars are coordinates of a coset space $\frac GH$. Equations of motion and Bianchi identities transform under the   isometry  group $G$, acting as dualities on the vectors as shown in \cite{Gaillard:1981rj}.  The isotropy group $H$ is its maximal compact subgroup.
Pure $\cN<4$ supergravities have no scalars.

{\footnotesize
\begin{table}[ht]
\begin{center}
\label{topotable}
\begin{tabular}{|c||c|c|c|c|c|c| }
\hline $\cN$ &  Duality group $G$ & isotropy $H$ & ${\cal M}_{scalar}$ & Spin-0 & Spin-1/2 & Spin-1  \\
\hline \hline
\hline $4$  &   $\SU(1,1)\otimes \SO(6)$ & $\U(4) $ &
$\frac{\SU(1,1) \otimes \SO(6) }{\U(1)\otimes \SO(6)}  $ & $1+1$& $4$ & 6 \\
\hline $5$  &  $\SU(1,5)$ &$\U(5)$  & $\frac{\SU(1,5)}
{S(\U(1)\times \U(5))}$ & $5+5$& $1 + 10$ & 10 \\
\hline $6$  &  $\SO^\star(12)$ &$\U(6)$ &
$\frac{\SO^\star(12)}{\U(1)\times \SU(6)}$ & $15+15$& $6+20$ & $1 + 15$ \\
\hline $8$&  $E_{7(7)}$ & $\SU(8)$ &
$\frac{ E_{7(7)} }{\SU(8)}$ & 70 & 56 & 28 \\
\hline
\end{tabular}
\end{center}
\caption{\it Scalar manifolds of $\cN \geq 4$ pure (no matter)  supergravities. We denote the scalar, spinor and vector field contents in each supergravity.
In some entries the $H$-representation is reducible.
Note that the $\mathbf{10}$ is the three-index antisymmetric representation of $\SU(5)$ and the $\mathbf{15}$ is the two-index antisymmetric representation of $\SU(6)$. In addition to the fields listed in the table, each supergravity contains a graviton and $\cN$ gravitini. }\label{tbl:G/H}
\end{table}
}

\subsection{Universal structure of on-shell linearized chiral superfields in \texorpdfstring{$\cN\geq 0$}{N greater equal 0} supergravity}
Linearized superfields contain the asymptotic physical states of the theory.  States of helicity $h$ are described by symmetric rank $|2h|$ spinors.
For every $\cN \geq 0$ there is a chiral superfield
$\bar{C}_{\dot\a\dot\b\dot\g\dot\delta}(y,\theta)$, whose lowest component in the $\theta$ expansion is the Weyl tensor
$\bar{C}_{\dot\a\dot\b\dot\g\dot\delta}(y)$ and describes a graviton of helicity $+2$.
For each $\cN$-extended  supergravity ($7 \geq \cN \geq 1$), there is a second chiral superfield, $\bar{\psi}_{\dot\a\dot\b\dot\g}^8$, $\bar{M}_{\dot\a\dot\b}^{78}$, $\bar{\chi}_{\dot\a}^{678}$, $W$, $W_{\a}$,  $W_{\a\b}$, $W_{\a\b\g}$, respectively.\footnote{The numerical subscripts 8,7,6, $\dots$ indicate the directions in space that are dropped in the descent from 8 to 7 to 6, etc.  See appendix~\ref{app:linchirSG}.}  Needless to say, the conjugates of these superfields are anti-chiral.
All of these superfields are singlets of the $\SU(\cN)$ global symmetry group. Derivations of their properties and details of their $\theta $ expansions can be found in appendix~\ref{app:linchirSG}, where it is also shown that there are no other chiral linearized superfields.
The analogous but simpler structure of ${\cal N}=4,3,2,1,0$  super-Yang-Mills theories  is illustrated in appendix~\ref{ss:N4321}.
\begin{table}[htbp]
\begin{center}

\begin{tabular}{|c||c|c|c|}
\hline
$\cN$ & Chiral & Anti-chiral & Helicity $h$,  $\U(1)_c$ weight  $c$\\
\hline \hline \hline
$\cN = 8$ &  $\bar{C}_{\dot\a \dot\b \dot\g \dot\d}(y, \theta) $ & $C_{\a\b\g\d}(\bar{y}, \bar{\theta})$  & $\pm 2, \, \, -$ \\
\hline
$\cN = 7$ & \begin{tabular}{@{}c@{}}  $\bar{C}_{\dot\a \dot\b \dot\g \dot\d}(y, \theta)$ \\ $\bar\psi_{\dot\a\dot\b \dot\g}^8(y, \theta)$ \end{tabular}  & \begin{tabular}{@{}c@{}}  $C_{\a\b\g\d}(\bar{y}, \bar{\theta})$ \\ $\psi_{\a\b\g 8}(\bar{y}, \bar\theta)$ \end{tabular} & \begin{tabular}{@{}c@{}} $\pm 2, \, \, 0$ \\ $\pm \frac{3}{2}, \, \, \pm \frac{7}{2}$ \end{tabular}  \\
\hline
$\cN = 6$ &  \begin{tabular}{@{}c@{}} $\bar{C}_{\dot\a \dot\b \dot\g \dot\d}(y, \theta)$ \\  $\bar{M}_{\dot\a\dot\b}^{78}(y, \theta)$ \end{tabular}   & \begin{tabular}{@{}c@{}} $C_{\a\b\g\d}(\bar{y}, \bar{\theta})$ \\ $M_{\a\b 78}(\bar{y}, \bar{\theta})$ \end{tabular} & \begin{tabular}{@{}c@{}} $\pm 2, \, \, 0$ \\ $ \pm 1, \, \, \pm 3$ \end{tabular} \\
\hline
$\cN = 5$ & \begin{tabular}{@{}c@{}} $\bar{C}_{\dot\a \dot\b \dot\g \dot\d}(y, \theta)$ \\ $\bar\chi^{678}_{\dot\a}(y, \theta)$ \end{tabular} & \begin{tabular}{@{}c@{}} $C_{\a \b \g \d}(\bar{y}, \bar{\theta} )$ \\ $ \chi_{\a 678}(\bar{y}, \bar\theta)$ \end{tabular} & \begin{tabular}{@{}c@{}} $\pm 2, \, \, 0$ \\ $\pm \frac{1}{2}, \, \, \pm \frac{5}{2}$ \end{tabular}  \\
\hline
$\cN = 4$ & \begin{tabular}{ @{}c@{}  }  $\bar{C}_{\dot\a \dot\b \dot\g \dot\d}(y, \theta)$ \\  $W(y, \theta) = \phi^{5678}(y, \theta)$ \end{tabular} & \begin{tabular} {@{}c@{} }   $C_{\a \b \g \d}(\bar{y}, \bar{\theta})$  \\ $\overline{W}(\bar{y}, \bar{\theta}) = \phi_{5678}(\bar{y}, \bar\theta) $  \end{tabular} &  \begin{tabular} {@{}c@{} } $ \pm 2, \, \, 0$  \\ $0, \,\, \pm 2$ \end{tabular}   \\
\hline
 $\cN = 3$ &  \begin{tabular}{ @{}c@{}  }  $\bar{C}_{\dot\a \dot\b \dot\g \dot\d}(y, \theta)$ \\  $W_{\a}(y, \theta) = \chi_{ \a  123}(y, \theta)$ \end{tabular}  &  \begin{tabular}{ @{}c@{}  }  $C_{\a \b \g \d}(\bar{y}, \bar\theta)$ \\  $\overline{W}_{\dot\a}(\bar{y}, \bar\theta) = \bar\chi^{123}_{\dot\a}(\bar{y}, \bar\theta)$ \end{tabular}  & \begin{tabular}{@{}c@{}}   $\pm 2, \, \, 0$  \\  $\mp \frac{1}{2}, \, \, \pm \frac{3}{2}$ \end{tabular}  \\
 \hline
$\cN = 2$  &  \begin{tabular}{ @{}c@{}  }  $\bar{C}_{\dot\a \dot\b \dot\g \dot\d}(y, \theta)$ \\  $W_{\a\b}(y, \theta) = M_{ \a\b  12}(y, \theta)$ \end{tabular}  &  \begin{tabular}{ @{}c@{}  }  $C_{\a \b \g \d}(\bar{y}, \bar\theta)$ \\  $\overline{W}_{\dot\a \dot\b}(\bar{y}, \bar\theta) = \bar{M}^{12}_{\dot\a \dot\b}(\bar{y}, \bar\theta)$ \end{tabular}  & \begin{tabular}{@{}c@{}}   $\pm 2, \, \, 0$  \\  $\mp 1, \, \, \pm 1$ \end{tabular}  \\
 \hline
 $\cN = 1$ &  \begin{tabular}{ @{}c@{}  }  $\bar{C}_{\dot\a \dot\b \dot\g \dot\d}(y, \theta)$ \\  $W_{\a \b \g}(y, \theta) = \psi_{ \a \b \g  1}(y, \theta)$ \end{tabular}  & \begin{tabular}{ @{}c@{}  }  $C_{\a \b \g \d}(\bar{y}, \bar\theta)$ \\  $\overline{W}_{\dot\a \dot\b \dot\g}(\bar{y}, \bar\theta) = \bar\psi^{1}_{\dot\a \dot\b \dot\g}(\bar{y}, \bar\theta)$ \end{tabular}  & \begin{tabular}{@{}c@{}}   $\pm 2, \, \, 0$  \\  $\mp \frac{3}{2}, \, \, \pm \frac{1}{2}$ \end{tabular} \\
 \hline
 $\cN = 0$ &  $\bar{C}_{\dot\a \dot\b \dot\g \dot\d}(x)$  & $C_{\a \b \g \d}(x)$  & $\pm 2, \, \, 0$ \\
 \hline

\end{tabular}
\end{center}
\caption{\it Chiral and anti-chiral on-shell multiplets for ${\cal N}=8$ to 0. The last column provides the helicity of the multiplet, where upper and lower sign corresponds to the chiral and anti-chiral one, respectively. The chiral $\U(1)_c$ weights are also given, which satisfy the simple relation $|h-c|=2$. For ${\cal N}=8$ the $c$-weights are not defined. Dimension is equal to the absolute value of helicity, $\Delta = |h|$, for all entries in this table.  }\label{tbl:chiralN80}
\end{table}

\subsection{\texorpdfstring{$\U(1)$}{U(1)} chirality/helicity in linearized superfields}
\label{ss:U1ch}
The  two global symmetries $\U(1)_{c}$ and $\U(1)_h$ discussed in section~\ref{review} for  linearized $\cN=4$ supergravity are also present for other values of $\cN$. Charges of the $\U(1)_c$ symmetry are those of the $\U(1)$ factor in the $R$-symmetry group $\U({\cal N})=\SU({\cal N})\times \U(1)$. For both symmetries $\theta^i_\alpha $ has charge 1/2, and $\bar \theta _{\dot \alpha }^i$ has charge $-1/2$, so that  supersymmetry generators change the charges of component fields by 1/2.
Furthermore the $\U(1)_{h}$ and $\U(1)_c$ act as phase transformations, and thus complex conjugate fields should have opposite charges.

For self-conjugate multiplets, the on-shell ${\cal N}=4$ super-Yang--Mills multiplet~(\ref{N4fieldsYM}) and the on-shell ${\cal N}=8$ supergravity multiplet (\ref{contentN8}), the last two requirements above fix the weights of all component fields. In particular, the middle (scalar) fields must have weight~0. The symmetry defined by these charges will be indicated as $\U(1)_h$, since fields have $\U(1)_h$ charge equal to their helicity.

For each $\cN<8$ there are two chiral superfields. Neither contains components in conjugate pairs, so the requirement of opposite charges for conjugate fields is not applicable. We define the $\U(1)_h$ charge of  each component as its helicity and assign the $\U(1)_c$ charge as its charge under the $\U(1)$ normal subgroup of the isotropy group $H$ of the scalar manifold. The latter
acts as a chiral transformation on fermions and by duality on vector fields, while the graviton is inert.
For $\cN=8$ the isotropy group, $\SU(8)$, is simple and does not contain a normal $\U(1)$ subgroup, which is consistent with the argument above that no other charges than those in $\U(1)_h$ can be defined for this multiplet.

In linearized $\cN$-supergravity, we have the following chiral superfields,
\begin{align}
\bar{C}_{\dot\a\dot\b\dot\g\dot\d}(y,\theta) \ (\cN\geq0),\quad \Phi_{\cN} (y,\theta)\ (7\geq \cN\geq1),
\end{align}
where $\Phi_{\cN}$ denotes the second chiral superfield for each $\cN$, as listed in table~\ref{tbl:chiralN80}. Their $\U(1)_c$ and $\U(1)_h$ transformation laws are
\begin{align}
&\U(1)_c:\quad \theta^i_\a\to \rme^{\frac{1}{2}\rmi a} \theta^i_\a,\quad \bar{C}_{\dot\a\dot\b\dot\g\dot\d}(y,\theta)\to \bar{C}_{\dot\a\dot\b\dot\g\dot\d}(y,\theta),\quad \Phi_{\cN}(y,\theta)\to \rme^{\frac{\rmi\cN a}{2}}\Phi_\cN(y,\theta), \nonumber\\
%\end{align}
%\begin{align}
&\U(1)_h:\quad \theta^i_\a\to   \rme^{\frac{1}{2}\rmi b} \theta^i_\a,\quad \bar{C}_{\dot\a\dot\b\dot\g\dot\d}(y,\theta)\to \rme^{2\rmi b}\bar{C}_{\dot\a\dot\b\dot\g\dot\d}(y,\theta),\quad \Phi_{\cN}(y,\theta)\to \rme^{\rmi\frac{\cN-4}{2}b}\Phi_\cN(y,\theta),
\end{align}
where $a$ and $b$ are constant parameters.  Information on these superfields and their $U(1)_{c,h}$ charges is given in table \ref{tbl:chiralN80}. The weights of the superfields are defined as the weights of their first components. In the 4-point linearized supersymmetric invariants studied in the next two sections of this paper,  the $\U(1)_c$ and $\U(1)_h$ anomalies for ${\cal N}<8$ are correlated.  Any given invariant is either anomalous under both symmetries or non-anomalous under both.

\section{Chiral superinvariants in \texorpdfstring{$\cN\geq 5$}{N >=5} supergravity and anomaly candidates}
\label{Cinv}
There are no matter multiplets in $\cN\geq 5$ supergravity, and this simplifies our analysis.
In particular all vertices carry the factor $1/\kappa^2$ and all propagators have the factor $\kappa^2$. Thus $L$-loop structures always contain the even power $\kappa^{2(L-1)}$, as first proposed in \cite{Kallosh:1980fi}.

\subsection{\texorpdfstring{${\cal N}=5$}{N=5}}
\label{ss:N5invariants}
We now list the  possible chiral non-invariant 4-point Lagrangian structures for $\cN=5$ supergravity.  We include factors of $\kappa$ so that  Lagrangians have dimension +4 and their contributions to the action
$\int \rmd^4 x \, {\cal L}(x)$ are dimensionless. We focus on candidate anomalous Lagrangians constructed from
the available chiral superfields, $\bar{C}_{\dot\a\dot\b\dot\g\dot\delta}$ and $\bar{\chi}^{\dot\alpha}= \bar{\chi}_{678}^{\dot\a}$.  (Of course, their conjugates can also be used.)

Since $\bar{\chi}_{\dot\a}$ is a fermionic superfield, only even powers can occur by Lorentz invariance.  Thus there are only three possible terms:
\begin{eqnarray}
{\cal L}_1^{2n}(x)&=&\kappa^{3+2n} \int \rmd^{10}\theta,  \partial^{2n} (\bar{\chi}_{\dot\a}\bar{\chi}^{\dot\a}  (x, \theta))^2  \sim  \partial^{2n} (\bar{\chi}\bar{\chi}(x) \partial C\partial C(x))+\cdots, \\
{\cal L}_2^{2n}(x)&=& \kappa^{6+2n} \int \rmd^{10}\theta  \, \partial^{2n} (\bar{\chi}_{\dot\a}\bar{\chi}^{\dot\a} \bar{C}_{\dot \b\dot \g\dot \delta\dot \eta}\bar C^{\dot \b\dot \g\dot \delta\dot \eta} (x, \theta) ) \sim \partial^{2n} (\bar{C}\bar{C} \partial C\partial C(x))+\cdots, \\
{\cal L}_3^{2n}(x)&=& \kappa^{9+2n}\int \rmd^{10}\theta \,  \partial^{2n} (\bar C_{\dot \b\dot \g\dot \delta\dot \eta}\bar C^{\dot \b\dot \g\dot \delta\dot \eta}(x, \theta))^2 \sim \partial^{2n} (\bar{C}\bar{C} \partial^4\bar{\chi}\partial^4\bar{\chi}(x))+\cdots\,.
\end{eqnarray}

For positive $n$, the notation  $\partial^{2n}(\cdots) $ indicates an arrangement
of spacetime derivatives acting on various superfields.  As $n$ increases the power of $\kappa$ needed to form a dimensionless contribution to the action also increases, and so does the loop order of the candidate invariant.  We use negative $n$ to schematically indicate a non-local expression (such as $1/stu$ in momentum space) and this decreases the relevant loop order. Since both $\bar{\chi}_{\dot \a}$  and $\bar{C}_{\dot\a\dot\b\dot\g\dot\d}$ carry only dotted indices, an even number of spacetime derivatives is required by Lorentz invariance.

Note that $\cL_1$ and $\cL_3$ have non-vanishing chirality weight, $5$ and $-5$ respectively, and a non-vanishing helicity weight, $-3$ and $+3$ respectively.  These potential  anomalous amplitudes are ruled out because  they contain odd powers of $\kappa$  which cannot occur in pure supergravity.

The invariant $\cL_2$ has chiral and helicity weight zero, and is therefore non-anomalous.  It contains the allowed MHV process $\<h^{- -}h^{- -} h^{+ +}h^{+ +}\>$ and its SUSY partners.  So $\cL_2$
 remains a possible candidate for counterterms starting at the 4-loop level. The explicit 4-loop amplitude  computations in \cite{Bern:2014sna} show that this divergence is absent.  This fact  remains unexplained.

Thus in $\cN=5$ supergravity all potential $\U(1)$ anomalous candidates are eliminated since they have odd dimension. We will see that this conclusion agrees with analysis of the double-copy construction.

\subsection{\texorpdfstring{$\cN=6$}{N=6}}
We proceed as above and construct  the possible chiral invariant actions that are quartic in the superfields $\bar{C}_{\dot \a \dot \b \dot \g \dot\delta}$ and $\bar{M}_{\dot\a\dot\b}= \bar{M}_{\dot\a\dot\b}{}^{78}$. There are three candidates, namely
\begin{eqnarray} \label{L1}
{\cal L}_1(x)^{2n}&=& \kappa^{6+2n}\int \rmd^{12}\theta \, \partial^{2n} (\bar{M}_{\dot\a\dot\b} \bar{M}^{\dot\a\dot\b}(x, \theta))^2\sim \partial^{2n}\, (\bar{M}\bar{M}\partial^2 C\partial^2 C)(x)+\cdots,\\ \label{L2}
{\cal L}_2(x)^{2n} &=&  \kappa^{8+2n} \int \rmd^{12}\theta \, \partial^{2n}(\bar M_{\dot\a\dot\b}\bar M^{\dot\a\dot\b}\bar{C}_{\dot\g\dot\delta\dot\eta\dot\epsilon}\bar{C}^{\dot\g\dot\delta\dot\eta\dot\epsilon} (x, \theta)) \sim  \partial^{2n}\, (\bar{C}\bar{C} \partial^2C\partial^2 C(x))+\cdots, \\
\label{L3}
{\cal L}_3(x)^{2n} &=&\kappa^{10+2n}\int \rmd^{12}\theta  \, \partial^{2n}\, (\bar{C}_{\dot\b\dot\g\dot\delta\dot\eta}\bar{C}^{\dot\b\dot\g\dot\delta\dot\eta} (x, \theta))^2\sim \, \partial^{2n} (\bar{C}\bar{C} \partial^4M\partial^4M)+\cdots.
\end{eqnarray}
Unlike the situation in $\cN=5$, all the three terms have even mass dimension and are possible anomalous amplitudes and candidates for  UV divergences. Their chirality weights are $6, 0, -6$ respectively, and they carry helicity $-2, 0, +2$.  Thus $\cL_2$ is chiral invariant and non-anomalous as was the case for $\cL_2$ in $\cN=5$.

Since $\cL_1^{2n}$ has the same helicity/chirality structure and dimension as the conjugate of $\cL_3^{2n-4}$, we focus on $\cL_1^{2n}$. For $n=0$ this  is the candidate 4-loop anomalous counterterm
\begin{equation}
{\cal M}_{\cN=6}^{\rm 4-loop} =  \kappa^6\int \rmd^4x\,\rmd^{12} \theta\, \bar M_{\dot \alpha \dot \beta }\bar M^{\dot \alpha \dot \beta }\bar M_{\dot \gamma \dot \delta }\bar M^{\dot \gamma \dot \delta }=
4 \kappa^6 \int \rmd^4x\, \rmd^{12} \theta\, \bar M_{\dot \alpha \dot \beta }\bar M^{ \dot \beta\dot \gamma  }\bar M_{\dot \gamma \dot \delta }\bar M^{\dot \delta\dot \alpha  },
 \label{AmplitudeN6}
\end{equation}
where the equality of the two forms is due to $\bar{M}_{\dot\alpha\dot\beta }\bar{M}^{\dot\gamma  \dot\beta}=\ft12\delta _{\dot\alpha} ^{\dot\gamma} \bar{M}_{\dot\delta\dot\epsilon }\bar{M}^{\dot\delta\dot\epsilon}$.

We are interested in the $vvhh$ amplitudes contained in the superspace invariant~\rf{AmplitudeN6}.  The first step is to substitute the $\theta$ expansion~\rf{M78D6}
of $\bar M^{78}$ truncated to the Maxwell and Weyl terms
\begin{eqnarray}
 \bar{M}_{\dot{\a}\dot{\b}}(x, \theta)= \bar{M}_{\dot{\a}\dot{\b}}(x) +\dots
-\ft1{6!}\theta^\a_i\theta^\b_j\theta^\g_k\theta^\delta_\ell\theta^\eta_m\theta^\sigma_n\varepsilon^{ijk\ell mn}\partial_{\a\dot{\a}}\partial_{\b\dot\b} C_{\g\delta\eta\sigma}(x).
 \label{superfield}
\end{eqnarray}
After performing the $\theta$ integration we find the candidate amplitude
\bea
  A_{vvhh}^{\rm 4-loop}&= \kappa^6\int \rmd^4x& \Big [ \bar M_{\dot \alpha \dot \beta } \bar M^{\dot \alpha \dot \beta }\partial_{\a}{}^{\dot{\gamma }}\partial_{\b}{}^{\dot\delta }  C_{\g\delta\eta\sigma}
  \partial^{(\a}{}_{\dot{\gamma }}\partial^{\b}{}_{\dot\delta } C^{\g\delta\eta\sigma)} \cr
  &&+ \bar M_{\dot \alpha \dot \beta }  \bar M^{\dot \gamma \dot \delta }\partial_{\a}{}^{\dot{\alpha }}\partial_{\b}{}^{\dot\beta } C_{\g\delta\eta\sigma}
  \partial^{(\a}{}_{\dot{\gamma }}\partial^{\b}{}_{\dot\delta } C^{\g\delta\eta\sigma)} \cr
%  \cr
  &&+ \bar M_{\dot \alpha \dot \beta }  \bar M^{\dot \gamma \dot \delta }
  \partial^{\a}{}_{\dot{\gamma }}\partial^{\b}{}_{\dot\delta } C^{\g\delta\eta\sigma} \partial_{(\a}{}^{\dot{\alpha }}\partial_{\b}{}^{\dot\beta } C_{\g\delta\eta\sigma)} \Big].
 \label{Avvhh}
\eea

This information may be recast in the language of amplitudes using the on-shell multi-spinor fields  (see
\cite{Kallosh:2008mq,Drummond:2010fp,Freedman:2011uc}). Using the amplitude relations of appendix~\ref{app:conventions},
  we obtain from the spinor contractions in \eqref{Avvhh} the supersymmetric helicity violating local 4-loop candidate counterterm:
\begin{equation}
 A_{vvhh}^{\rm 4-loop}= \kappa^6 \achirlam{12}^2\chirlam{34}^4(s^2+t^2+u^2)\delta ^4(P).
  \label{Avvgg3}
\end{equation}
The corresponding
helicity violating non-local 1-loop candidate amplitude is obtained by dividing by $stu$:
\begin{equation}
 A_{vvhh}^{\rm 1-loop}= \frac{(s^2+t^2+u^2) }{stu}\achirlam{12}^2\chirlam{34}^4\delta ^4(P).
 \label{Avvgg1loop}
\end{equation}

To obtain the candidate superamplitude, we use the superwavefunction for any particle at the position $I$:
\begin{eqnarray}
  \Omega _I&=& v_I^- + \eta_I^i f_i^- + \ft12\eta_I^i \eta_I^j  s_{ij}+\ft1{3!} \eta_I ^i\eta_I ^j\eta_I ^k f_{ijk}^+ +\ft1{4!2} \eta_I ^i\eta_I ^j\eta_I ^k\eta_I ^\ell\varepsilon _{ijk\ell mn} v^{mn+} \nonumber\\
  &&+ \ft1{5!}\eta_I ^i\eta_I ^j\eta_I ^k\eta_I ^\ell\eta_I ^m \varepsilon _{ijk\ell mn}
  \Psi ^{n+}+\ft1{6!} \eta_I^i\eta_I^j\eta_I ^k\eta_I ^\ell\eta_I ^m\eta_I ^n \varepsilon _{ijk\ell mn}h^{++}\,.
 \label{superampl}
\end{eqnarray}
We can then define the 4-point  superamplitude that has \eqref{Avvgg3} as a component. Using (\ref{sIJ}) and (\ref{equality4pts})
\begin{align}
\cM_4^{\rm 4-loop}=\kappa^6 \delta ^{12}(Q)(s^2+t^2+u^2)\frac{\achirlam{12}^2}{\chirlam{34}^2}\delta ^4(P),
 \label{Asuper}
\end{align}
and one shows that the candidate 4-loop amplitude is permutation invariant. The 1-loop candidate superamplitude is
\begin{align}
\cM_4^{\rm 1-loop}= \delta ^{12}(Q)\frac{s^2+t^2+u^2}{stu}\frac{\achirlam{12}^2}{\chirlam{34}^2}\delta ^4(P)\,.
\label{super1}\end{align}

We can reproduce the $vvhh$ amplitude~\eqref{Avvgg1loop} from the candidate superamplitude~(\ref{super1}) by applying  sixth-order $\eta_3$ and $\eta_4$ derivatives to project out  two gravitons.
Using
\begin{equation}
  \left(\frac{\partial }{\partial \eta _I}\right)^6= \prod_{i=1}^6 \frac{\partial }{\partial \eta _I^i}\,,
 \label{defdeta6}
\end{equation}
we have on the one hand
\begin{equation}
  \left(\frac{\partial }{\partial \eta _I}\right)^6\Omega _I = h^{++}\,,
 \label{deta6Omega}
\end{equation}
and on the other hand
\begin{equation}
 \left(\frac{\partial }{\partial \eta _3}\right)^6\left(\frac{\partial }{\partial \eta _4}\right)^6 \delta ^{12}(Q)= \chirlam{34}^6.
 \label{resultsup}
\end{equation}
We then obtain
\begin{align}
A_{vvhh}^{\rm 1-loop} &=   \chirlam{v^-v^- h^{++} h^{++}} \nonumber \\ &= \left(\frac{\partial }{\partial \eta _3}\right)^6\left(\frac{\partial }{\partial \eta _4}\right)^6 \cM_4^{\rm 1-loop} \nonumber \\ &=  \frac{s^2+t^2+u^2}{stu}\achirlam{12}^2\chirlam{34}^4\delta ^4(P).
 \label{Avvgg4}
\end{align}
As expected, permutation invariance of the candidate superamplitude is reduced to the Bose symmetries  $1\leftrightarrow 2$ and $3\leftrightarrow 4$.

In summary, in $\cN=6$ supergravity there are anomalous candidate superamplitudes that describe both potential 4-loop UV divergence and the 1-loop processes.
In section~\ref{N6doubcop} we will reproduce the 1-loop structure~\eqref{Avvgg4} using the double-copy method. In section~\ref{N6doubcopDan} we will show, however, that these candidate anomalous superamplitudes are not present in the effective action of $\cN=6$ supergravity.

\subsection{ \texorpdfstring{$\cN= 8$}{N=8} }\label{N8action}
We now discuss possible  $\U(1)_h$ anomalies in linearized $\cN=8$ supergravity. As explained in section \ref{ss:U1ch}, $\U(1)_c$ weights are not defined in this case,
 but we should still consider possible $\U(1)_h$ anomalies.
We focus on the 4-point invariant with lowest mass dimension:
\be
\int \rmd^{16}\theta(\bar{C}_{\dot\a\dot\b\dot\g\dot\delta}\bar{C}^{\dot\a\dot\b\dot\g\dot\delta}(x, \theta))^2\sim \bar{C}\bar{C}\partial^4 C\partial^4C(x)+\cdots.
\ee
We have written only the 4-graviton coupling on the right side and omitted other terms.
It is now easy to see that this is actually $\U(1)_h$ invariant: the  weight of the chiral measure is $-8$ whereas the helicity  weight of the superfields in the integrand is $+8$. Thus, as expected, there are no chiral/helicity anomalies in $\cN= 8$.

\section{Double copy SYM and anomaly candidates in  \texorpdfstring{$\cN\geq 5$}{N>=5} supergravity}
\label{DC}
\subsection{ \texorpdfstring{$\cN= 5$}{N=5} supergravity and \texorpdfstring{$\cN=4\otimes \cN=1$}{N=4 x N=1} SYM }
$\cN=5$ supergravity amplitudes are expressed by the double copy $\cN=4\otimes \cN=1$ SYM. As shown in section~\ref{ss:N5invariants}, there is no possible candidate for the anomalous amplitudes in $\cN=5$ supergravity. Let us consider the possible anomalous amplitudes from the double copy viewpoint. Since $\cN=4$ pure SYM has no helicity violating amplitudes, we focus on $\cN=1$ SYM.%, which has no self-conjugate structure.

In $\cN=1$ pure SYM, we have the following 4-point interactions,
\begin{align}
\cL_1=& \int \rmd^4x\,\rmd^2\theta\, \chi^\a \chi_\a \chi^\b \chi_\b,\\
\cL_2=&\int \rmd^4x\,\rmd^2\theta\,\chi^\a \chi_\a \bar{F}_{\dot{\b}\dot{\g}}\bar{F}^{\dot{\b}\dot{\g}},\\
\cL_3=&\int \rmd^4x\,\rmd^2\theta\, \bar{F}_{\dot{\a}\dot{\b}}\bar{F}^{\dot{\a}\dot{\b}} \bar{F}_{\dot{\g}\dot{\delta}}\bar{F}^{\dot{\g}\dot{\delta}}.
\end{align}
Spacetime derivatives can be added, but only in even numbers.
Since the theory has no dimensional parameter, $\cL_{1,3}$, which have odd  dimension, cannot appear in the one-loop effective action. Then, only $\cL_2$ is a candidate, but this term is a self-conjugate, chiral invariant, i.e. not $\U(1)$ anomalous:
\begin{align}
\cL_2=\int \rmd^4x\,\rmd^2\theta\, \chi^\a \chi_\a \bar{F}_{\dot{\b}\dot{\g}}\bar{F}^{\dot{\b}\dot{\g}}\sim F_{\a\b}F^{\a\b}\bar{F}_{\dot{\g}\dot{\d}}\bar{F}^{\dot{\g}\dot{\d}}+\cdots.
\end{align}
Therefore
 $\cN=1$ SYM does not have anomalous 4-point amplitudes in any loop order. Interestingly, the reason for the absence of anomalous amplitudes in $\cN=1$ SYM is almost the same as that in $\cN=5$ supergravity. Thus, the absence of anomalous amplitudes in $\cN=5$ supergravity is confirmed from the double copy viewpoint.

\subsection{ \texorpdfstring{$\cN= 6$}{N=6} supergravity and \texorpdfstring{$\cN=4 \otimes\cN=2$}{N=4 x N=2} SYM }\label{N6doubcop}
 In this section, we consider the $\cN=6$ supergravity amplitude from the double copy viewpoint. The $\cN=6$ supergravity amplitudes are given by the double copy $\cN=4\ \text{SYM} \otimes \cN=2\ \text{SYM}$. It is known that $\cN=4$ SYM has no helicity violating amplitudes, so we discuss such amplitudes in $\cN=2$ SYM.

We study the candidate helicity violating one-loop invariant
\begin{align}\label{Interaction}
& \delta ^4(P)\int\rmd^4\theta\, W(p_1) W(p_2) W(p_3) W(p_4) \frac{s^2+t^2+u^2}{stu} \nonumber
\\ & \sim \delta ^4(P)\phi(p_1) \phi(p_2) F^{\alpha \beta} (p_3) F_{\alpha \beta} (p_4)  \frac{s^2+t^2+u^2}{stu}  +\cdots.
\end{align}
It has full permutation symmetry, which will provide a permutation symmetry of the double-copy gravitational amplitudes, as in examples shown in  \cite{Carrasco:2013ypa}. Note the nonlocal interaction which makes this structure dimensionless.

Our invariant contains the $\phi\phi vv$ process on the second line of \eqref{Interaction}.
To express this as an on-shell amplitude we need the $\theta$ expansion (conjugate of (\ref{barWN2}))
\begin{equation}\label{N2superfield}
W = \phi + \theta^\alpha_ i \varepsilon ^{ij} \psi_{\alpha j } + \ft12\theta_i^\alpha \theta_j ^\beta \varepsilon ^{ij} F_{\alpha \beta },
\end{equation}
and the on-shell fields
%To translate this interaction into the corresponding amplitude, we rewrite the component fields as
\begin{align}
F_{\alpha\beta} (p) &= \lambda_{\alpha}(p) \lambda_{\beta}(p) \hat{v}^+, \nonumber\\
\psi_{\alpha}(p) &= \lambda_{\alpha}(p) \hat{\psi}, \nonumber\\
\phi &= \hat{\phi}.
\end{align}
From the first term in the second line of~\eqref{Interaction}, we can read off $\langle\hat\phi \hat\phi  v^+ v^+ \rangle$ as
\begin{equation}
\langle 34 \rangle^2 \frac{s^2+t^2+u^2}{stu}.
\end{equation}
One finds that this amplitude is one of the components of the superamplitude
\begin{equation}
\mathcal{A}_4^{\mathcal{N} = 2 \text{ anomalous}} =  \delta^4(  {Q} )  \frac{s^2+t^2+u^2}{stu}.\label{anomalous2}
\end{equation}
We use the double copy formula  (note $t= 2p_2\cdot p_3$)
\begin{align}
\mathcal{A}_{4}^{\mathcal{N} = 6 \text{ anomalous}}  \propto  s\, t  \, \delta^4 (P) \mathcal{A}_4^{\mathcal{N} = 4  \text{tree}} \times \mathcal{A}_4^{\mathcal{N} = 2 \text{ anomalous}}.
\end{align}
The $\cN= 4$ tree-level 4-point superamplitude is given by
\begin{equation}\label{anomN2}
\mathcal{A}_4^{\mathcal{N} = 4  \text{tree}} = \frac{\delta^8 ( {Q  ) } }{\langle 12 \rangle \langle 23 \rangle \langle 34 \rangle \langle 41 \rangle }.
\end{equation}
Combining \eqref{anomalous2} and \eqref{anomN2}, we find
\begin{align}
\mathcal{A}_{4}^{\mathcal{N} = 6 \text{ anomalous}}  & \propto \frac{  s\, t \, }{ \langle 12 \rangle \langle 23 \rangle \langle 34 \rangle \langle 41 \rangle  } \, \frac{s^2+t^2+u^2}{stu} \delta^{12} ({Q})\delta^4 (P) \nonumber
\\& \propto  \frac{[12]^2  }{ \langle 34 \rangle^2}  \frac{s^2+t^2+u^2}{stu} \delta^{12} ({Q}) \delta^4 (P),
\end{align}
in full agreement with  \rf{super1}. Thus we have found a supersymmetric candidate for a $\U(1)$ anomalous structure in $\cN=6$ supergravity  via two very different constructions.

\subsection{No 1-loop anomalous amplitudes in \texorpdfstring{$\cN=2$}{N=2} SYM theory} \label{N6doubcopDan}

In this section we examine the basic $\cN=2$ SYM field theory and show explicitly that such anomalous amplitudes are forbidden at one loop order.
As described in  \cite{Carrasco:2013ypa},  the construction of loop amplitudes and candidate counterterms in $\cN\geq 4$ supergravity theories proceeds by the double copy method.  This method begins with the observation that the spectrum of particle states in the $\cN$-extended supergravity theory is isomorphic to the direct product of states in two SYM theories,  specifically  $\cN=4\otimes (\cN -4)$ SYM.  We  are concerned with anomalous 4-point amplitudes associated with a possible chiral anomaly of the $\U(1)_R$ current.
Since the $\cN=4$ SYM theory does not contain  $\U(1)_R$, the lowest order anomalous supergravity amplitude requires a  non-vanishing 1-loop SYM amplitude on the $(\cN-4)$ side with non-conserved $\U(1)_R$ charge.  We now show rather simply that such amplitudes vanish in $\cN=2$ SYM theory.

In the previous section we constructed the linearized chiral $\cN=2$ SYM superinvariant~(\ref{Interaction}).  The 4th order terms in the  $\theta$ expansion  of $W(x,\theta)^4$ describe the 4-point processes  $\phi^2 F_{\a\b}F^{\a\b}$, $(\psi_{\a i} \psi^{\a i})^2$, $\phi\psi_{\a i}\psi_{\b j} \varepsilon^{ij} F^{\a\b}$.  The corresponding superamplitude is proportional to $\d^{(4)}(Q)$ with no further factors of $\eta^i_I$. Thus,  given any one of the three amplitudes above, the remaining two are determined by SUSY Ward identities.  Therefore we focus on the simplest amplitude, namely $\phi^2 F_{\a\b}F^{\a\b}$, and we consider the corresponding 1-loop Feynman diagrams.

For this we use the $\cN=1$ description of the minimal $\cN=2$ theory presented in (6.57)-(6.59) of  \cite{Freedman:2012zz}.
This model contains the bosonic fields $A_\mu^\ada,~\phi^\ada$, $\bar\phi^\ada$, and the Majorana fermions\footnote{The $P_L$ and $P_R$ projections of a Majorana field are identified as  Weyl spinors, e.g. $P_L\lambda_\a= \lambda_\a$ and $P_R\lambda_{\dot\a}= \lambda_{\dot\a}.$ Thus $\bar\chi P_L\lambda = \chi^{\a}\l_\a$.} $\lambda^\ada$, $\chi^\ada$, where $\ada,\adb,\adc$ are indices of the adjoint representation of a compact gauge group with structure constants $f^{\ada\adb\adc}$.  The structure of $\cN=2$ SUSY is hidden in this notation, but is realized by the assignment $\psi_1^\ada = \lambda^\ada$, $\psi_2^\ada=\chi^\ada$.
\begin{figure}[ht]
\centering
\begin{subfigure}[b]{0.3\textwidth}
\begin{center}
\begin{tikzpicture}[scale=1.5]
\draw[dashed](-0.5,0) -- (0,0);
\draw (0,0) -- (1,0);
\draw (0,0) -- (0,-1);
\draw (0,-1) -- (1,-1);
\draw (1,0) -- (1,-1);
\draw[dashed] (1,0) -- (1.5, 0);
\draw[photon] (-0.5,-1) -- (0, -1);
\draw[photon] (1,-1) -- (1.5,-1);
\node[label=above:$x$] at (0,0) {};
\node[label=above:$y$] at (1,0) {};
\node[label=below:$v$] at (0,-1) {};
\node[label=below:$u$] at (1,-1) {};
\end{tikzpicture}
\end{center}
\caption{}
\label{fig:subfigb}
\end{subfigure}
~
\begin{subfigure}[b]{0.3\textwidth}
\begin{center}
\begin{tikzpicture}[scale=1.5]
\draw[dashed](-0.5,0) -- (0,0);
\draw (0,0) -- (1,0);
\draw (0,0) -- (0,-1);
\draw (0,-1) -- (1,-1);
\draw (1,0) -- (1,-1);
\draw[photon] (1,0) -- (1.5,0);
\draw[photon] (-0.5,-1) -- (0, -1);
\draw[dashed] (1,-1) -- (1.5,-1);;
\node[label=above:$x$] at (0,0) {};
\node[label=above:$u$] at (1,0) {};
\node[label=below:$v$] at (0,-1) {};
\node[label=below:$y$] at (1,-1) {};
\end{tikzpicture}
\end{center}
\caption{}
\label{fig:subfigc}
\end{subfigure}
~
\begin{subfigure}[b]{0.3\textwidth}
\begin{center}
\begin{tikzpicture}[scale=1.5]
\draw[dashed] (-0.5,0) -- (0,0);
\draw[dashed] (0,0) -- (-0,-1);
\draw[photon] (-0.5,-1) -- (0,-1);
\draw[photon] (0,0) -- (1,0);
\draw[dashed] (1,0) -- (1.5,0);
\draw[dashed] (0,-1) -- (1,-1);
\draw[dashed] (1,0) -- (1,-1);
\draw[photon] (1,-1) -- (1.5,-1);
\node[label=above:$\phi$] at (-0.2,-0.1) {};
\node[label=below:$\bar{\phi}$] at (-0.2,0.1) {};
\node[label=above:$\phi$] at (-0.2,-1.1) {};
\node[label=below:$\bar{\phi}$] at (0.1,-0.9) {};
\node[label=below:$\phi$] at (0.9,-0.9) {};
\node[label=above:$\bar{\phi}$] at (1.2,-1.1) {};
\node[label=below:$\bar{\phi}$] at (1.2,0.1) {};
\node[label=above:$\phi$] at (1.2,-0.1) {};
\node[label=right:$\longleftarrow$] at (0.9, -0.5) {};
\end{tikzpicture}
\end{center}
\caption{}
\label{fig:subfiga}
\end{subfigure}
\caption{One boson and two fermion diagrams are shown. The letters $x$, $y$, $u$, $v$ indicate spacetime points in (\subref{fig:subfigb}) and (\subref{fig:subfigc}). In (\subref{fig:subfiga}), $\phi$, $\bar{\phi}$ are fields at the vertices, and the arrow indicates the vanishing Wick contraction. }
\label{N=2processes}
\end{figure}
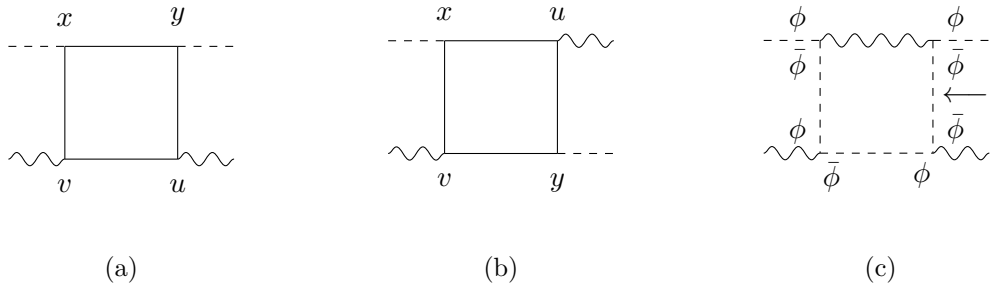

The interaction terms of this theory consist of the usual non-abelian gauge interactions for the adjoint representation plus the Yukawa and quartic terms:
\be \label{N2int}
L_{\rm int} = -\sqrt2 f^{\ada\adb\adc}[\bar\chi^\ada (\phi^\adb P_R +\bar \phi^\adb P_L)\lambda^\adc] +\ft12 f^{\ada\adb\adc}f^{\ada\add\ade}(\bar \phi^\adb\phi^\adc) (\bar\phi^\add\phi^\ade)\,.
\ee
  An  $\cN=1$ chiral superfield
$\phi(x,\theta) = \phi(x) + \theta^\a\chi_\a(z) + \theta^\a\theta_\a F(x)$ carries the multiplet $\U(1)_R$ charge $r_\phi$, and $\theta_\a$ is assigned $r_\theta=1$.  The component charges  are then $r_\phi$, $r_\chi=r_\phi-1$, $r_F=r_\phi-2.$  \,In the gauge multiplet, the real quantities $A_\mu,~F_{\mu\nu}$ are $\U(1)_R$ inert, and the gaugino $\l_\a$ carries $r_\l=1$. The Yukawa interaction  in \rf{N2int} is invariant if we choose $r_\phi=2$.
The entire action is then $\U(1)_R$ invariant as an $\cN=1$ theory.  With the assignments   $\psi_1^\ada = \lambda^\ada$, $\psi_2^\ada=\chi^\ada$,  this extends to the $\U(1)_R$ of \,$\cN=2$ SUSY.
This symmetry prohibits the process $\phi^2 F_{\a\b}F^{\a\b}$.

All Feynman diagrams with two external $\phi$ fields and two external gluons vanish because the additive $\U(1)_R$ charge is conserved at all interaction vertices.  We now confirm this at 1-loop order.
Two fermion diagrams and a representative boson loop diagram are indicated in figure~\ref{N=2processes}.
 We indicate  Wick contractions along the top line of the first diagram with gauge indices suppressed.
\be\label{wick}
\int \rmd^4x\,\bar\phi(x)  (\bar\chi P_L\bar \phi\lambda)_x \int \rmd^4y (\bar\lambda P_L \bar\phi \chi)_y\,\bar\phi(y).
\ee
The Wick contraction $\<\l(x)\bar\l(y)\>\sim \dslash\frac{1}{(x-y)^2}$ gives the massless fermion propagator in spacetime.  But there are adjacent $P_L$ projectors, so we find the structure
\be\label{vanish1}
P_L  \dslash_x\frac{1}{(x-y)^2}P_L =0.
\ee
It vanishes by elementary chirality.  The same argument applies to the case where two $\chi$ fields are contracted.  The second diagram in the figure is a little more complicated, but it also vanishes because there are an odd number of $\gamma$-matrices between the projectors:
\be\label{vanish2}
P_L \dslash_x\gamma _\mu \dslash_uP_L = 0.
\ee
It is simpler to analyze bosonic diagrams.  Since there are two external $\phi$-fields, one encounters the vanishing  Wick contraction $\<\phi \phi\>=0$ along the loop and the amplitude vanishes.  Note that our arguments apply to the process $\phi+\phi \to n$-gluons, and the amplitude vanishes for all gluon helicities.\footnote{We thank Lance Dixon for this observation.}

\subsection{ \texorpdfstring{$\cN= 8$}{N=8} supergravity and \texorpdfstring{$\cN=4 \otimes\cN=4$}{N=4 x N=4} SYM }
We consider the anomalous amplitudes in $\cN=8$ supergravity from the double copy $\cN=4 \otimes\cN=4$ SYM. In $\cN=4$ SYM, we have only one chiral invariant action with four superfields,
\begin{align}
\cL=\int \rmd^4x\,\rmd^{8}\theta\, (\bar{F}_{\dot{\a}\dot{\b}} \bar{F}^{\dot{\a}\dot{\b}})^2\sim \bar{F}_{\dot{\a}\dot{\b}} \bar{F}^{\dot{\a}\dot{\b}} \partial^2F_{\a\b}\partial^2F^{\a\b}+\cdots.
\end{align}
This is the helicity preserving 4-point interaction, which of course does not give anomalous amplitudes. In this maximal supersymmetric YM, there are no other helicity violating superinvariants as in the maximal supergravity case shown in section~\ref{N8action}. Therefore, we conclude that $\cN=8$ supergravity does not have anomalous amplitudes from not only the superinvariant analysis in section~\ref{N8action}, but also the double copy viewpoint.

\section{  Subtleties in Anomalies:  the  \texorpdfstring{$\eps/\eps$}{epsilon/epsilon} effect   }

This section is based on discussions of amplitude anomalies with Z. Bern, L. Dixon and R. Roiban.
We begin with a comment that provides the background for this section.
The candidate chirality violating amplitudes studied in this paper would represent an anomaly of the global $\U(1)_c$ symmetry.  In theories with $\U(1)$ chiral anomalies, invariance holds at the classical level but fails at 1-loop order.  When dimensional regularization to $d=4-2\eps$ is used,  the difficulty arises because
one must supply a prescription for the dimensional continuation of $\gamma^5= \rmi \gamma^0\gamma^1\gamma^2\gamma^3$,  see for  example~\cite{tHooft:1972tcz,Peskin:1995ev}.  The result is that the numerators of 1-loop Feynman diagrams vanish linearly as $\eps\rightarrow 0$,  but the loop integration can produce a $1/\eps$ pole.  The net result is a finite anomalous amplitude.  If the pole  is absent there is no anomaly at least in the amplitude under study.

The argument in \cite{Carrasco:2013ypa} in $\cN=4$ supergravity and  in sections~\ref{Cinv} and~\ref{DC} for $\cN=6$ is based on the properties of linearized chirality violating supersymmetry invariants,
 their associated superamplitudes and the simple double copy relations that are appropriate in this context.
We note the following concerning the invariants (\ref{M1})-(\ref{M2}) in the $\cN=4$ theory.  The first one is non-local, but the second is \emph{local}.  This means that  (\ref{M2}) with an extra $1/\eps$ factor is a candidate 1-loop counterterm in the effective action and this fact is reflected in the local superamplitude (\ref{III}).  It is clearly correlated with the order $\eps/\eps$ rational terms in 1-loop $\cN=0$ gauge theory amplitudes.  In contrast we note, looking at \rf{L1}-\rf{L3}, that there is no \emph{local} candidate chirality violating 1-loop invariant in $\cN=6$ supergravity.  More to the point,  the chirality violating 1-loop invariant \rf{Interaction} in $\cN=2$ SYM is non-local, and there is no local dimensionless structure which would contain the $1/\eps$ pole.  This proves that the $\eps/\eps$ effect is  absent in  $\cN=2$ SYM and motivates  the unregulated 1-loop calculations of section~\ref{N6doubcopDan}, which confirm this.

For further justification, we now relate our approach to that of \cite{Carrasco:2012ca}, in which 1-loop supergravity and SYM  amplitudes are calculated using unitarity cuts and dimensional continuation from $d=6$  to $d=4-2\eps$. Anomalous rational terms may appear due to the $\eps/\eps$  effect.  The issue of concern to us is that $\cN=4$ supergravity coupled to two vector multiplets can be realized as a double copy in two ways, namely as $(\cN=4)\times (\cN=0)$ and $(\cN=2)\times(\cN=2)$.  The first construction clearly gives anomalous amplitudes from the $\eps/\eps$ effect in the $(\cN=0)$ factor. In section~3 of \cite{Carrasco:2012ca}  it is shown that both constructions yield identical amplitudes for all $d$ and for all helicity configurations.
This appears to indicate
a 1-loop anomaly in $\cN=2$ SYM, which would contradict our result.  On the other hand, the calculations in section~4 of \cite{Carrasco:2012ca} confirm that 1-loop chirality violating amplitudes vanish as $\eps\to 0.$  The explanation of this apparent paradox is that the double-copy calculations are performed in the integrands of unitarity cuts, which contain the ${\cal O}(\epsilon)$ terms of the $\cN=2$ gauge theory.  However this zero does not survive the final integration needed to produce the anomalous amplitudes of $\cN=4$ supergravity plus two vector multiplets.

\section{Summary and discussion}

This work was motivated by the results of \cite{Carrasco:2013ypa} in $\cN=4$ supergravity.  In that paper two independent linearized supersymmetric chiral invariants that contain $U(1)_c$ anomalous amplitudes were found, and subsequent 4-loop  calculations in \cite{Bern:2013uka}  showed that these structures are actually present as UV divergences.  Therefore we undertook a study of the potential anomalous chiral invariants in $\cN=5,6,8$ supergravity, confining our attention to quartic invariants that contain 4-point processes. As a preliminary step we classified the linearized chiral superfields
in all supergravities.  There is one such superfield in the self-conjugate $\cN=8$ theory and two for each $\cN$ in the range $7\geq \cN> 0.$  The analogous classification in super-Yang-Mills theories for $4\geq \cN > 0$ was also given, since we constructed potential invariants by the double copy method as well as intrinsically in supergravity.

The $\U(1)_c$ symmetry is  the normal $\U(1)$ subgroup of the  $R$-symmetry group $\U(\cN)$ for $7\geq \cN> 0.$   Since the group $\SU(8)$ is simple, $\U(1)_c$ does not exist for $\cN=8$. However, we define $\U(1)_h$ which counts the helicity of asymptotic physical and exists for all $\cN$.  The charges of the two symmetries $\U(1)_{c,h}$ are correlated, so that any amplitude is either anomalous or non-anomalous under both. It was shown in \cite{Marcus:1985yy} that there is an $R$-symmetry anomaly in $\cN=4$ supergravity, but not for $\cN\geq 5$.  An $R$-symmetry anomaly is a failure of the conservation law for the $R$-current in 3-point functions in external backgrounds.
One might suspect that this fact is also relevant for the question of anomalous contributions to the 4-particle S-matrix.

Let us come to the main results.
For $\cN=5$ supergravity there are three  quartic chiral superspace invariants, two of which violate chirality.  However the physical dimension of these two is odd, and they cannot be constructed from the Feynman rules of extended supergravity because they would violate elementary dimensional analysis.  For $\cN =6$  there are again three chiral invariants, and two violate chirality. These invariants have even physical dimension, so a more detailed argument is needed to investigate whether they actually appear.  For this purpose we derive the superamplitude that corresponds to the invariants by two methods, first by working directly from the candidate invariants and second using their construction via the double copy method from the product $\cN=4\ \text{SYM} \otimes \cN=2\ \text{SYM}$.
Both methods lead to the same chirality violating superamplitude. The double copy method works initially at the 1-loop level and yields a non-local structure which is then promoted to a local form which is a candidate 4-loop UV divergence.  We then present a simple argument based on the $\cN=2$ SYM Lagrangian,  that there are no anomalous 1-loop amplitudes.  Finally we arrive at  $\cN=8$ supergravity where there is only one chiral superspace invariant which is in fact helicity conserving.

In summary there are anomalous amplitudes in $\cN=4$ supergravity.  There are candidate anomalous chiral superspace invariants in $\cN =5,~6$ supergravity, but their contributions actually vanish when the structure of the theories is examined. In $\cN=8$ there are no anomalous candidates.  This strengthens one's confidence that the presence or absence of the $R$-symmetry anomaly is significant.
\

\

\

\noindent{\bf {Acknowledgments:}} We are grateful to Z. Bern, J.J. Carrasco, L. Dixon, R. Roiban  for stimulating discussions.  The work  of DZF, RK, DM, and YY is supported by SITP and by the US National Science Foundation grant PHY-1316699. The research of DZF is partly supported by the US National Science Foundation grant PHY-1620045 and as a Templeton Visiting Professor at Stanford. The work of AVP is supported in part by the Interuniversity Attraction Poles Programme
initiated by the Belgian Science Policy (P7/37), and in part by
support from the KU Leuven C1 grant ZKD1118 C16/16/005. AVP also thanks the Department of Physics of Stanford University for hospitality during a visit in which part of this work was performed.

\appendix

\section{Conventions}
\label{app:conventions}
We use complex conjugation that does not change the order of fermions, and that acts on the fermion components as
\begin{equation}
  (\psi _\alpha )^* = \bar \psi _{\dot \alpha }\,,\qquad (\psi ^\alpha )^* = - \bar \psi ^{\dot \alpha }\,.
 \label{complexfinal}
\end{equation}
A real fermion bilinear is rewritten as follows
\begin{equation}
  \bar \chi \psi= \chi ^\alpha \psi _\alpha + \bar \chi _{\dot \alpha  }\bar \psi ^{\dot \alpha }\,,\qquad \bar \chi \gamma ^\mu \psi  = \chi ^\alpha(\gamma ^\mu )_{\alpha \dot \alpha } \bar \psi^{\dot \alpha } + \bar \chi _{\dot \alpha  }(\gamma ^\mu) ^{\dot \alpha \alpha }\psi _{ \alpha }
 \label{fermbil}
\end{equation}
where $(\gamma ^\mu )_{\alpha \dot \alpha }=(\sigma ^\mu)_{\alpha \dot \alpha }$ and $(\gamma ^\mu )^{\dot \alpha  \alpha }=(\bar \sigma ^\mu)^{\alpha \dot \alpha }$, as given in \cite{Freedman:2012zz} in (2.2).

We write the algebra of covariant derivatives for ${\cal N}$-extended supersymmetry as
\begin{equation}
  D^i_{\alpha} \bar D_{j\dot \alpha }+ \bar D_{j\dot \alpha }D^i_{\alpha}= \delta ^i_j\partial _{\alpha \dot \alpha }\,,\qquad \partial _{\alpha \dot \alpha }\equiv -2(\gamma ^\mu )_{\alpha \dot \alpha }\partial _\mu \,.
 \label{covDalg}
\end{equation}
This leads to $\partial _{\alpha \dot \alpha }\partial ^{\alpha \dot \alpha }= 8\partial _\mu \partial ^\mu $ or
$\partial _{\alpha \dot \alpha }\phi \partial ^{\alpha \dot \alpha }\phi '= 8\partial _\mu \phi \partial ^\mu \phi '$.
Complex conjugation also raises or lowers the index $i$, and $ (D^i_{\alpha} )^* = \bar D_{i\dot \alpha }$.

We often use the chiral basis where
\begin{equation}
 y^{\alpha \dot \alpha }= x^{\alpha \dot \alpha }+ \ft12\bar \theta ^{i\dot \alpha }\theta _i^\alpha\,,\qquad  \bar D_{i\dot \alpha }=-\frac{\partial }{\partial \bar \theta ^{i\dot \alpha }}\,,\qquad D^i_{\alpha }= \frac{\partial }{\partial \theta _i^{\alpha }}-\bar \theta ^{i\dot \alpha }\partial _{\alpha \dot \alpha }\,,
 \label{chiralDbasis}
\end{equation}
or the anti-chiral one
\begin{equation}
 \bar y^{\alpha \dot \alpha }= x^{\alpha \dot \alpha }- \ft12\bar \theta ^{i\dot \alpha }\theta _i^\alpha\,,\qquad  \bar D_{i\dot \alpha }=-\frac{\partial }{\partial \bar \theta ^{i\dot \alpha }}+\ft12 \theta _i^\alpha \partial _{\alpha \dot \alpha }\,,\qquad D^i_{\alpha }= \frac{\partial }{\partial \theta _i^{\alpha }}\,.
 \label{achiralDbasis}
\end{equation}

The momenta of particle $I$ are written in terms of two-component commuting spinors as \cite{Bianchi:2008pu,Elvang:2015rqa}
\begin{equation}
  p_{I\alpha \dot \alpha }= -2(\gamma ^\mu )_{\alpha \dot \alpha }p_{I\mu }= 2\lambda _{I\alpha }\tilde {\lambda} _{I\dot \alpha }\,.
 \label{ptolambda}
\end{equation}
Particle helicities are expressed in terms of these spinors as e.g.
\begin{equation}
  \overline M_{\dot \alpha \dot \beta }(p) =\bar \lambda _{\dot \alpha }(p)\bar \lambda _{\dot \beta }(p)v^-\,,\qquad C_{\alpha \beta \gamma \delta }(p)= \lambda _\alpha(p) \lambda _\beta(p) \lambda _\gamma(p) \lambda _\delta(p) h^{++}\,.
 \label{FlambdaC}
\end{equation}
The contraction of these spinors for $p_I$ and $p_J$ is then indicated as
\begin{equation}
  \chirlam{IJ}=-\chirlam{JI} = \lambda _I^\alpha \lambda_{J\alpha} \,,\qquad \achirlam{IJ}=-\achirlam{JI}=(\chirlam{IJ})^*= \bar \lambda _{I\dot \alpha }\bar \lambda_J^{\dot \alpha }\,.
 \label{defchirlam}
\end{equation}
They satisfy the relations
\begin{equation}
  \chirlam{IJ}\achirlam{JI}= 2p_I\cdot p_J= s_{IJ}\,,\qquad \sum_J \chirlam{IJ}\achirlam{JK}=0\,,
 \label{sIJ}
\end{equation}
for lightlike momenta and $s_{IJ}= (p_I+p_J)^2$. With these one shows that for 4-point functions
\be
\frac{\achirlam{12}^2}{\chirlam{34}^2}= \frac{\achirlam{13}^2}{\chirlam{24}^2}= \frac{\achirlam{14}^2}{\chirlam{32}^2}= \frac{\achirlam{23}^2}{\chirlam{14}^2}= \frac{\achirlam{24}^2}{\chirlam{13}^2}= \frac{\achirlam{34}^2}{\chirlam{12}^2}\,.
\label{equality4pts}
\ee
We use the shortcuts $s=s_{12}=s_{34}$, $t=s_{23}=s_{14}$ and $u=s_{13}=s_{24}$.

When using super-wave functions, the supersymmetry generators $Q_\alpha ^i$ are represented as
\begin{equation}
 Q_\alpha ^i=\sum_I \lambda _{I\alpha } \eta _I^i \,,
 \label{defQ}
\end{equation}
with anticommuting vectors $\eta_I^i$. The $\int \rmd^{2{\cal N}}\theta $ is then effectively replaced by the $\delta $ function
\begin{equation}
  \delta ^{2{\cal N}}(Q) = \prod_{i=1}^{{\cal N}} \prod_{\alpha =1}^2 \delta (Q_\alpha ^i)= \prod_{i=1}^{{\cal N}} \prod_{\alpha =1}^2\sum_I \lambda _{I\alpha } \eta _I^i = \prod_{i=1}^{{\cal N}} \sum_{I>J}  \chirlam{IJ}\eta _I^i\eta _J^i\,.
 \label{del2NQ}
\end{equation}

\section{Linearized chiral superfields in SYM}\label{ss:N4321}
\subsection{Universal structure of on-shell linearized chiral superfields in SYM}
As discussed in section~\ref{ss:pureExtSG}, $\cN\geq0$ supergravity has a universal structure. Here we show that $\cN\geq0$ SYM also has the same structure, that is, singlet fields under $\SU(\cN$) are the lowest component of (anti-)chiral superfields, and the $\theta^{\cN}$ component of any chiral superfields is the lowest component of another anti-chiral superfield.

\subsection {\texorpdfstring{$\cN= 4$}{N=4} YM}
We begin with the field content of $\cN = 4$ SYM. The on-shell degrees of freedom are given by
\begin{equation}
\{\cN=4\ \text{SYM}\}=\{F_{\a\b}, \lambda_{\a i}, \phi_{ij}, \bar{\lambda}_{\dot\a}^i, \bar{F}_{\dot\a \dot\b}\} .
 \label{N4fieldsYM}
\end{equation}
In our notation, lower flavor indices correspond to the $\mathbf{4}$ of $\SU(4)$ and upper flavor indices correspond to the $\bar{\mathbf{4}}$. Note that there is a duality constraint given by
\begin{equation}
\phi_{ij} = \frac{1}{2} \varepsilon_{ijkl} \phi^{kl}\,,\qquad \phi^{ij} = \phi_{ij}^*\,.
  \label{phiijN4}
\end{equation}
In a suitable basis, we can define the Bianchi identities for the fields in the multiplet, which define also themselves superfields
\begin{align}
D^i_{\a} \phi_{jk} &= 2 \d^i_{[j} \lambda_{\a k]}, \label{bianchi1} \\
D^i_{\a} \lambda _{\b k} &= \d^i_k F_{\a\b}. \label{bianchi2}
\end{align}
These can be inverted to obtain
\begin{align}
\lambda_{\a i } &= \frac{1}{3} D^j_{\a} \phi_{j i},  \nonumber\\
F_{\a\b} &= \frac{1}{4} D^i_{\alpha} \lambda_{\beta i}.
\end{align}

Note that we can derive further identities:
\begin{align}
\bar{D}_{\dot\a i}\phi_{jk} &= \varepsilon_{ijk\ell}\bar{\lambda}^\ell _{\dot{\a}},\nonumber\\
\bar{D}_{\dot\a i} \lambda _{\a j} &= \partial_{\dot\a \a} \phi_{ij}, \nonumber\\
\bar{D}_{\dot\a i} F_{\a\b} &= \partial_{\dot\a (\a}\lambda  _{\b) i}, \label{bianchi4}
\end{align}
where we have used $\{ D^i_{\alpha}, \bar{D}_{\dot\a j}  \} = \d^i_j \partial_{\a \dot\a}$ to derive the second and third equations.

With Bianchi identities, we can show that $F_{\a\b}$ is an anti-chiral superfield as follows:
\begin{align}
D^m_\g D^\ell_\b D^i_\a \phi_{jk}=2\delta^{i\ell}_{jk}D_\g^mF_{\a\b},
\end{align}
where we have used \eqref{bianchi1}, \eqref{bianchi2} and $\delta $ with more indices is the antisymmetric product of $\delta $ symbols, e.g. $\delta _{ij}^{k\ell}=\delta _{[i}^k\delta _{j]}^\ell$. For $m\neq i,j,k,\ell$, the left hand side vanishes because $D^m_\g D^\ell_\b D^i_\a \phi_{jk}=D^\ell_\b D^i_\a (D^m_\g \phi_{jk})=0$ due to \eqref{bianchi1}. Choosing further for any $m$ a couple $[ij]=[k\ell]$ of different indices, we find
\begin{align}
D_\g^mF_{\a\b}=0.
\end{align}

On the chiral basis  (\ref{chiralDbasis}), we can expand the chiral superfield $\bar{F}_{\dot{\a}\dot{\b}}$ in $\cN=4$ YM as
\begin{align}\label{N4superfield}
\bar{F}_{\dot\a \dot\b}(y, \theta) &= \bar{F}_{\dot \a \dot\b}(y) + \theta^{\a}_i \partial_{\a (\dot\a}\bar{ \lambda}^i_{\dot\b)} + \frac{1}{2!}  \theta^{\a}_i \theta^{\b}_j \partial_{\a \dot\a} \partial_{\b \dot\b} \phi^{ji} \nonumber
\\& + \frac{1}{3!} \theta^{\a}_i \theta^{\b}_j \theta^{\gamma}_k \varepsilon ^{kji\ell} \partial_{\a \dot\a} \partial_{\b \dot\b} \lambda_{\gamma \ell}  - \frac{1}{4!}  \theta^{\a}_i \theta^{\b}_j \theta^{\gamma}_k \theta^{\delta}_\ell \varepsilon ^{\ell kji} \partial_{\a \dot\a} \partial_{\b \dot\b} F_{\gamma \delta}.
\end{align}
To derive this expansion, we have used the identities (\ref{bianchi1})-(\ref{bianchi2}) and (\ref{bianchi4}).

\subsection{\texorpdfstring{$\cN<4$}{N<4} chiral superfields from \texorpdfstring{$\cN=4$}{N=4} SYM}
We derive $\cN<4$ superfields from the truncation of $\cN=4$ SYM. First let us start with $\cN=3$ SYM, which is obtained by setting $\theta_4=0$. As is the case of $\cN=7$ supergravity, which will be discussed in section~\ref{N<8}, this truncation does not reduce the degrees of freedom in the $\cN=4$ theory. One can confirm it by performing the truncation of the superfield~\eqref{N4superfield}.

Next, we consider $\cN=2$ SYM by setting $\theta^\alpha _{3,4}=0$, which gives
\begin{align}\label{N2superfieldapp}
\bar{F}_{\dot\a \dot\b}(y, \theta) &= \bar{F}_{\dot \a \dot\b}(y) + \theta^{\a}_i \partial_{\a (\dot\a}\bar{ \lambda}^j_{\dot\b)} + \frac{1}{2!}  \theta^{\a}_i \theta^{\b}_j \varepsilon^{ji}\partial_{\a \dot\a} \partial_{\b \dot\b} \phi_{34},
\end{align}
where $i,j=1,2$. The last component $\phi_{34}\equiv\bar\phi$ is a singlet scalar field. On-shell degrees of freedom in $\cN=2$ SYM are summarized as
\begin{align}
\{\cN=2\text{ SYM}\}=\{F_{\a\b},\lambda^i_\a, \phi\}+\cc.
\end{align}
Note also that from $\cN=4$ SYM we obtain
\begin{equation}
  \phi_{12} = \phi^{34}=\phi.
 \label{dualityN2}
\end{equation}
From \eqref{bianchi1}, one can immediately find that the scalar superfield $\bar\phi(x,\theta,\bar{\theta})$ is an anti-chiral superfield. Its component expansion in the anti-chiral basis  (\ref{achiralDbasis}) is
 \begin{align}
\overline W(\bar y,\theta)= \bar{\phi}(\bar y,\theta)=\bar{\phi}(\bar y)-\bar{\theta}^{i\dot\a}\varepsilon_{ij} \bar{\lambda}^j_{\dot\a}+\frac{1}{2}\varepsilon_{ij}\bar{\theta}^{i\dot\a} \bar{\theta}^{j\dot\b} F_{\dot\a\dot\b}\,.\label{barWN2}
 \end{align}

We impose the condition $\theta_{2,3,4}=0$ to obtain $\cN=1$ SYM. The curvature superfield is
\begin{align}
\bar{F}_{\dot\a \dot\b}(x, \theta) &= \bar{F}_{\dot \a \dot\b}(x) + \theta^{\a}\partial_{\a (\dot\a}\bar{ \lambda}_{\dot\b)},
\end{align}
where we have omitted the index $i=1$ of $\bar{\lambda}^i_{\dot \alpha }$. \eqref{bianchi2} shows that $\bar{\lambda}_{\dot\a}(x,\theta,\bar{\theta})$ is an anti-chiral superfield, whose component expression is
\begin{align}
\overline{W}_{\dot \alpha }(\bar{y},\bar{\theta})=\bar{\lambda}_{\dot\a}(\bar{y},\bar{\theta})=\bar{\lambda}_{\dot\a}(x)-\bar{\theta}^{\dot\b}\bar{F}_{\dot\a \dot\b}.
\end{align}

We summarize the (anti-)chiral superfields of $\cN\geq0$ SYM in table~\ref{tbl:NSYM}.
\begin{table}[htbp]
\begin{center}
\begin{tabular}{|c|c|c|c|}
\hline
$\mathcal{N}$ & Chiral & Anti-chiral & Helicity \\
\hline
$\mathcal{N} = 4$ &  $\bar{F}_{\dot\a\dot\b}(y, \theta) $ & $F_{\a\b}(\bar{y}, \bar{\theta})$  & $\pm 1$ \\
\hline
$\cN = 3$ & \begin{tabular}{@{}c@{}}  $\bar{F}_{\dot\a\dot\b}(y, \theta)$ \\ $\bar\chi_{\dot\a}^4(y, \theta)$ \end{tabular}  & \begin{tabular}{@{}c@{}}  $F_{\a\b}(\bar{y}, \bar{\theta})$ \\ $\chi_{\a 4}(\bar{y}, \bar\theta)$ \end{tabular} & \begin{tabular}{@{}c@{}} $\pm 1$ \\ $\pm \frac{1}{2}$ \end{tabular}  \\
\hline
$\cN = 2$ &  \begin{tabular}{@{}c@{}} $\bar{F}_{\dot\a\dot\b}(y, \theta)$ \\  $W(y, \theta) = \phi^{34}(y, \theta)$ \end{tabular}   & \begin{tabular}{@{}c@{}} $F_{\a\b}(\bar{y}, \bar{\theta})$ \\ $\bar{W}(\bar{y}, \bar{\theta}) = \phi_{34}(\bar{y}, \bar{\theta}) $ \end{tabular} & \begin{tabular}{@{}c@{}} $\pm 1$ \\ $ 0$ \end{tabular} \\
\hline
$\cN = 1$ & \begin{tabular}{@{}c@{}} $\bar{F}_{\dot\a \dot\b}(y, \theta)$ \\ $W_{\a} = \lambda_{\a 1}(y, \theta)$
\end{tabular} & \begin{tabular}{@{}c@{}} $F_{\a \b}(\bar{y}, \bar\theta)$ \\ $\bar{W}_{\dot\a} = \bar \lambda^{1}_{\dot\a}(\bar{y}, \bar\theta)$
\end{tabular} & \begin{tabular}{@{}c@{}} $\pm 1$ \\ $\mp \frac{1}{2}$ \end{tabular}  \\
\hline
$\mathcal{N} = 0$ & $\bar{F}_{\dot\a \dot\b}(x)$ & $F_{\a \b}(x)$ & $\pm 1$ \\
\hline
\end{tabular}
\end{center}
\caption{\it Properties of linearized chiral/anti-chiral superfields in $\cN$-extended supersymmetric Yang-Mills theory. Dimension $\Delta$ is related to the absolute value of helicity  $\Delta = |h|+1$, for all entries in this table. }
\label{tbl:NSYM}
\end{table}

Finally, we show the proof of the following universality:  {\it In $\cN \geq 4$ SYM, the $\theta^{\cN}$ component of any chiral superfields is the lowest component of another anti-chiral superfield.}
In any $\cN$-SYM, there exists (at least) one anti-chiral superfield $F_{\a\b}$. Here we start with the case $\cN=4$ SYM. Using \eqref{bianchi1}, \eqref{bianchi2}, \eqref{bianchi4}, one can show the following equations:
\begin{align}
\bar{D}_{\ell\dot{\a}}F_{\a\b}=&\partial_{\dot{\a}(\a}\lambda _{\b)\ell},\label{1D}\\
\bar{D}_{k\dot{\b}}\bar{D}_{\ell\dot{\a}}F_{\a\b}=&\partial_{\dot{\a}(\a}\partial_{\dot{\b}\b)}\phi_{k\ell},\label{2D}\\
\bar{D}_{j\dot{\g}}\bar{D}_{k\dot{\b}}\bar{D}_{\ell\dot{\a}}F_{\a\b}=&\varepsilon_{k\ell ji}\partial_{\dot{\a}(\a} \partial_{\dot{\b}\b)} \bar{\lambda }^i_{\dot{\g}},\label{3D}\\
\bar{D}_{i\dot{\delta}}\bar{D}_{j\dot{\g}}\bar{D}_{k\dot{\b}}\bar{D}_{\ell\dot{\a}}F_{\a\b}=&\varepsilon_{k\ell ji}\partial_{\dot{\a}(\a}\partial_{\dot{\b}\b)}\bar{F}_{\dot{\delta}\dot{\g}}\label{4D}.
\end{align}
In $\cN=4$ SYM, the $\theta^{4}$ component is proportional to $\bar{D}_{i\dot{\delta}}\bar{D}_{j\dot{\g}}\bar{D}_{k\dot{\b}}\bar{D}_{\ell\dot{\a}}F_{\a\b}|_{\theta=0}$, which corresponds to $\varepsilon_{k\ell ji}\partial_{\dot{\a}(\a}\partial_{\dot{\b}\b)}\bar{F}_{\dot{\delta}\dot{\g}}$. This shows the universal structure in the $\cN=4$ case.

Next, let us (formally) consider the $\cN=3$ case, which is given by the truncation of $\cN=4$ SYM. The r.h.s. of \eqref{4D} always vanishes in $\cN=3$ since there are only three indices. This means
\begin{align}
\bar{D}_{i\dot{\delta}}\bar{D}_{j\dot{\g}}\bar{D}_{k\dot{\b}}\bar{D}_{\ell\dot{\a}}F_{\a\b}=\varepsilon_{k\ell j4}\partial_{\dot{\a}(\a} \partial_{\dot{\b}\b)}\bar{D}_{i\dot{\delta}} \bar{\lambda }^4_{\dot{\g}}=0,
\end{align}
where we have used \eqref{3D}. Thus, we have shown that $\bar{D}_{i\dot{\delta}}\bar{\chi}^4_{\dot{\g}}=0$, or equivalently, $\bar{\chi}^4_{\dot{\g}}$ is a chiral superfield. Note also that the l.h.s. of \eqref{3D} corresponds to the $\theta^{3}$ component of the anti-chiral superfield $F_{\a\b}$ in $\cN=3$ SYM. Then, the universality holds in the $\cN=3$ SYM case.  One can continue this procedure to the $\cN=1$ case in the same way. This proves the universal structure.

One can find and prove the same universal structure in supergravity as in the SYM case.

\section{Linearized chiral superfields in  supergravity}
\label{app:linchirSG}
\subsection{\texorpdfstring{$\cN=8$}{N=8} supergravity}
The on-shell degrees of freedom are given by
\begin{equation}
\left\{{\cal N}=8\right\}=  \left\{C_{\a\b\g\delta}, \ \psi_{\a\b\g i },\ M_{\a\b ij},\ \chi_{\a ijk}, \ \phi^{ijk\ell }, \ \bar{\chi}_{\dot{\a} }^{ijk}, \  \bar{M}_{\dot \a \dot \b}^{ ij }, \  \psi_{\dot\a \dot\b \dot\g i}, \ \bar{C}_{\dot\a \dot\b \dot\g \dot\delta}\right\}.
 \label{contentN8}
\end{equation}
The scalar fields satisfy
\begin{equation}
  (\phi _{ijk\ell })^* = \phi ^{ijk\ell }=\frac{1}{4!}\varepsilon ^{ijk\ell mnpq}\phi _{mnpq}\,.
 \label{realityphi}
\end{equation}

There are a number of Bianchi identities on the superfields that start with these fields as lowest components  \cite{Howe:1980th,Howe:1981gz}:
\begin{eqnarray}
D_\a^i\phi _{jk\ell m}&=&4\delta^i_{[j}\chi_{\a k\ell m]},\nonumber\\
D_\alpha ^i \chi _{\beta jk\ell }&=& 3 \delta ^i_{[j}M_{\alpha \beta k\ell ] },\nonumber\\
D_\alpha ^i M_{ \beta\gamma  jk }&=& 2\delta ^i_{[j}\psi _{\alpha \beta \gamma k] },\nonumber\\
D_\alpha ^i\psi _{\beta \gamma \delta j }&=&\delta ^i_j C_{\alpha \beta \gamma \delta },\label{cn1}
\end{eqnarray}
and its complex conjugates, e.g.
\begin{equation}
  \bar{D}_{i\dot{\a}}\phi ^{jk\ell m} =4\delta_i^{[j}\bar \chi^{ k\ell m]}_{\dot \alpha }.
 \label{cn2}
\end{equation}

These identities can be understood from the helicities and $\SU({\cN})$ content of the theory, written in (\ref{contentN8}). From the representation point of view other terms with different $\SU({\cal N})$ and helicity representations could appear in the right-hand side of the equations in (\ref{cn1}). However, there exists no on-shell degrees of freedom for these representations. The physics determines the structure of these equations. The exact equations actually define the superfields that appear in the right-hand side. They can be obtained from the inverses of (\ref{cn1}):
\begin{eqnarray}
 \chi _{\alpha ijk} & = & \ft15 D_\alpha ^\ell  \phi _{\ell ijk}, \nonumber\\
 M_{\alpha \beta ij} & = & \ft16 D_\alpha ^k \chi _{\beta kij},\nonumber\\
 \psi _{\alpha \beta \gamma i } &=& \ft17 D_\alpha ^jM_{\beta \gamma ji},\nonumber\\
 C_{\alpha \beta \gamma \delta }&=& \ft18 D_\alpha ^i \psi _{\beta \gamma \delta i}.
 \label{inverted}
\end{eqnarray}

Using the algebra $\{D_\alpha \bar D_{\dot \alpha }\}=\partial _{\alpha \dot \alpha }$, we can then also prove
\begin{eqnarray}
 \bar D_{i\dot \alpha }\chi _{\beta jk\ell } & = & \partial _{\dot \alpha\beta }\phi _{ijk\ell }\,, \nonumber\\
\bar D_{i\dot \alpha }M_{\beta \gamma jk}& = & \partial _{\dot \alpha (\beta  }\chi _{\gamma) ijk}\,,\nonumber\\
\bar D_{i\dot \alpha }\psi _{\beta\gamma \delta j} &=&\partial _{\dot \alpha (\beta  }M_{\gamma \delta) ij}\,,\nonumber\\
\bar D_{i\dot \alpha }C_{\beta\gamma \delta \epsilon }&=& \partial _{\dot \alpha (\beta  }\psi _{\gamma \delta \epsilon) i}\,.
 \label{barDchi}
\end{eqnarray}

\subsection{Chiral superfield in \texorpdfstring{$\cN=8$}{N=8} supergravity}

The above relations allow us to prove that $\bar C_{\dot \alpha \dot \beta \dot \gamma \dot \delta }$ is a chiral superfield, satisfying
\begin{equation}
  \bar D_{i\dot \epsilon }\bar C_{\dot \alpha \dot \beta \dot \gamma \dot \delta } = 0 \,.
 \label{chiralC}
\end{equation}
Indeed, the Bianchi identity~(\ref{cn1}) can be iterated to
\begin{equation}
D^i_{\a}  D^j_{\b}  D^k_{\g}  D^\ell_{\d} \phi _{mnpq}= 4! \delta ^{\ell kji}_{mnpq}C_{\alpha \beta \gamma \delta }\,.
 \label{valueCdel}
\end{equation}
If we act on this with a covariant derivative $D_\epsilon ^r$ with $r\neq i,j,k,\ell,m,n,p,q$, the first of (\ref{cn1}) implies
that $\delta ^{\ell kji}_{mnpq}D_\epsilon ^r C_{\alpha \beta \gamma \delta }=0$. For any $r$ we can find 4 different indices $[ijk\ell]=[mnpq]$ such that this proves the vanishing of $D_\epsilon ^r C_{\alpha \beta \gamma \delta }$. The complex conjugate implies that $\bar C_{\dot \alpha \dot \beta \dot \gamma \dot \delta }$ is chiral.

Using the same name for a superfield and its first component, we can then use (\ref{cn1}), (\ref{barDchi}) to identify the components of that chiral superfield. We obtain
\begin{align}
\bar{C}_{\dot\a\dot\b\dot\g\dot\d}(y, \theta)&= \bar C_{\dot\a\dot\b\dot\g\dot\d}(y) +\theta^{\a}_i  \partial_{\a (\dot{\a} } \bar{\psi}^i_{\dot \b \dot \g \dot \d)}
+\ft12\theta^{\a}_i \theta^{\b}_j \partial_{\a (\dot \a} \partial_{\b \dot\b} \bar{M}^{ji}_{\dot \g \dot\d)}
+\ft1{3!} \theta^{\a}_i \theta^{\b}_j \theta^{\g}_k \partial_{\a (\dot\a} \partial_{\b \dot\b}  \partial_{\g \dot \g} \bar{\chi}_{\dot \d)}^{kji} \nonumber \\
&+ \ft1{4!}\theta^{\a}_i \theta^{\b}_j \theta^{\g}_k \theta^{\d}_{\ell } \partial_{\a \dot\a} \partial_{\b \dot\b}  \partial_{\g \dot \g} \partial_{\d \dot \d} \phi^{\ell k ji}
+\ft1{5!3!} \theta^{\a}_i \theta^{\b}_j \theta^{\g}_k \theta^{\d}_{\ell } \theta^{\eps}_{m} \partial_{\a \dot\a} \partial_{\b \dot\b}  \partial_{\g \dot \g} \partial_{\d \dot \d} \chi_{\eps npq} \varepsilon ^{ijk\ell mnpq} \nonumber \\
&+\ft1{6!2} \theta^{\a}_i \theta^{\b}_j \theta^{\g}_k \theta^{\d}_{\ell } \theta^{\eps}_{m} \theta^{\zeta}_{n} \partial_{\a \dot \a} \partial_{\b \dot \b} \partial_{\g \dot \g} \partial_{\d \dot \d} M_{\eps \zeta pq} \varepsilon ^{ijk\ell mnpq} \nonumber\\
&+ \ft1{7!}  \theta^{\a}_i \theta^{\b}_j \theta^{\g}_k \theta^{\d}_{\ell } \theta^{\kappa _1}_{m} \theta^{\kappa _2}_{n} \theta^{\kappa _3}_{p}
\partial_{\a \dot \a} \partial_{\b \dot \b} \partial_{\g \dot \g} \partial_{\d \dot \d} \psi_{\kappa_1\kappa _2\kappa _3 q} \varepsilon ^{ijk\ell mnpq} \nonumber \\&
+\ft1{8!} \theta^{\a}_i \theta^{\b}_j \theta^{\g}_k \theta^{\d}_{\ell } \theta^{\kappa _1}_{m} \theta^{\kappa _2}_{n} \theta^{\kappa _3}_{p}\theta ^{\kappa _4}_q
  \partial_{\a \dot \a} \partial_{\b \dot \b} \partial_{\g \dot \g} \partial_{\d \dot \d} C_{\kappa_1\kappa _2\kappa _3\kappa _4} \varepsilon ^{ijk\ell mnpq}\,,
  \label{chiralbarCN8}
\end{align}
where the symmetrization of indices in the first line concerns only the dotted indices.
Since the (mass) dimension of $\theta $ is $\ft12$, the structure of the superfield is consistent with the dimensions of the fields as in
  table~\ref{tbl:dimfields}.
\begin{table}[H]
\begin{center}
 $ \begin{array}{|c|ccccc|}\hline
  \mbox{Field} & C_{\a\b\g\delta}&\psi_{\a\b\g  }& M_{\a\b }&\chi_{\a }&\phi  \\[2mm]
  \mbox{Dimension} & 2  & \ft32 & 1 & \ft12 & 0 \\[2mm] \hline
\end{array}$
 \caption{\it (Mass) dimensions of the fields. We do not indicate the $\SU({\cal N})$ labels since this table is valid for different ${\cal N}$ %with different representations.
 \label{tbl:dimfields}}
\end{center}
 \end{table}

\subsection{\texorpdfstring{$\cN<8$}{N<8} chiral superfields from \texorpdfstring{$\cN=8$}{N=8}}
\label{N<8}
In the following section, we will mainly focus on chiral superfields in $\cN=5,6$.
  It is known that $\cN=5,6$ supergravity is derived by truncation of $\cN=8$ supergravity. The structure of $\cN=8$ supergravity would be extremely useful. In particular, we can easily find chiral superfields in $\cN=5,6$ supergravity from $\cN=8$. But the procedure is more generally applicable for all ${\cal N}<8$.

The truncation procedure consists in putting some components of $\theta _i^\alpha $ to zero in (\ref{chiralbarCN8}) and its complex conjugate.\footnote{ The similar truncation is also done for superwave functions in ref.~\cite{Damgaard:2012fb}.} A first trivial example is when we try to truncate to ${\cal N}=7$. Then we would just put $\theta _8^\alpha =0$ in (\ref{chiralbarCN8}). The last term would not appear anymore, however the field $C_{\alpha \beta \gamma \delta }$ still appears in the anti-chiral superfield. Another example is that in the second term $\bar{\psi}^8_{(\dot \b \dot \g \dot \d)}$ does not appear anymore. However, it also appears in the complex conjugate. Equivalently, we can see this that the complex conjugate field $\psi_{\kappa_1\kappa _2\kappa _3 8}$ is still present in the one but last term. This shows in this formalism that ${\cal N}=7$ is not different from ${\cal N}=8$.

We now consider the truncation to ${\cal N}=6$. We put $\theta _8^\alpha =\theta _7^\alpha =0$. Now e.g. $\bar{\psi}^8_{(\dot \b \dot \g \dot \d)}$ does not enter in the second term, but neither does $\psi_{\kappa_1\kappa _2\kappa _3 8}$ in the one but last term, since for that the set $[ijk\ell mnp]$ in the Levi-Civita symbol should contain 7, which is then contracted with the vanishing $\theta _7^\alpha $. Considering in this way the full superfield, and restricting now the $\SU({\cal N})$ indices $i,\ldots $ to $\{1,\ldots ,6\}$, the chiral superfield  is
\begin{align}
\bar{C}_{\dot\a\dot\b\dot\g\dot\d}(y, \theta)&= \bar C_{\dot\a\dot\b\dot\g\dot\d}(y) +\theta^{\a}_i  \partial_{\a (\dot{\a} } \bar{\psi}^i_{\dot \b \dot \g \dot \d)}
+\ft12\theta^{\a}_i \theta^{\b}_j \partial_{\a (\dot \a} \partial_{\b \dot\b} \bar{M}^{ji}_{\dot \g \dot\d)}
+\ft1{3!} \theta^{\a}_i \theta^{\b}_j \theta^{\g}_k \partial_{\a (\dot\a} \partial_{\b \dot\b}  \partial_{\g \dot \g} \bar{\chi}_{\dot \d)}^{kji} \nonumber \\
&+ \ft1{4!}\theta^{\a}_i \theta^{\b}_j \theta^{\g}_k \theta^{\d}_{\ell } \partial_{\a \dot\a} \partial_{\b \dot\b}  \partial_{\g \dot \g} \partial_{\d \dot \d} \phi^{\ell k ji}
+\ft1{5!} \theta^{\a}_i \theta^{\b}_j \theta^{\g}_k \theta^{\d}_{\ell } \theta^{\eps}_{m} \partial_{\a \dot\a} \partial_{\b \dot\b}  \partial_{\g \dot \g} \partial_{\d \dot \d} \chi_{\eps n78} \varepsilon ^{ijk\ell mn} \nonumber \\
&+\ft1{6!} \theta^{\a}_i \theta^{\b}_j \theta^{\g}_k \theta^{\d}_{\ell } \theta^{\eps}_{m} \theta^{\zeta}_{n} \partial_{\a \dot \a} \partial_{\b \dot \b} \partial_{\g \dot \g} \partial_{\d \dot \d} M_{\eps \zeta 78} \varepsilon ^{ijk\ell mn} \,.
  \label{chiralbarCN6}
\end{align}
For investigations of ${\cal N}=6$ we may delete the $[78]$ indices. E.g. the spin 1 field $M_{\eps \zeta }=M_{\eps \zeta 78}$ is a singlet of $\SU(6)$. The scalar field $\phi^{ijk\ell}$ has 15 complex components, which can be written in terms of
\begin{equation}
  \phi_{ij}=\frac{1}{4}\varepsilon_{ijk\ell mn}\phi ^{k\ell mn}=\phi _{ij78}\,,\qquad \mbox{or}\qquad \phi ^{ijk\ell}= \frac{1}{2}\varepsilon^{ijk\ell mn}\phi _{mn}\,,
 \label{phi6cc}
\end{equation}
and we could have used this in the fifth term in (\ref{chiralbarCN6}).

The superfield~(\ref{chiralbarCN6}) is obviously still chiral since in the chiral basis~(\ref{chiralDbasis}) the chirality for ${\cal N}=8$ means that the superfield does not depend on the $\bar \theta$ and thus a fortiori not on those $\bar\theta ^i$ with $i\in\{1,\ldots ,6\}$. This shows that the components of the multiplet are
\begin{equation}
  \left\{{\cal N}=6\right\}= \left\{C_{\a\b\g\delta},\ \psi_{\a\b\g i},\  M_{\a\b ij},\ \chi_{\a}^{ijk},\ \phi^{ij}, \ \bar{\chi}_{\dot{\a}}^{i},\ \bar{M}_{\dot{\a}\dot{\b}}\right\}+ \cc\,,
\label{N6fields}
\end{equation}
where we indicated explicitly the complex conjugates of the fields in (\ref{chiralbarCN6}).

We can truncate the theory to $\cN = 5$ by taking $\theta_{6,7,8}^\alpha =0$ and consider the truncated superfield
\begin{align}
\bar{C}_{\dot\a\dot\b\dot\g\dot\d}(y, \theta)&= \bar C_{\dot\a\dot\b\dot\g\dot\d}(y) +\theta^{\a}_i  \partial_{\a (\dot{\a} } \bar{\psi}^i_{\dot \b \dot \g \dot \d)}
+\ft12\theta^{\a}_i \theta^{\b}_j \partial_{\a (\dot \a} \partial_{\b \dot\b} \bar{M}^{ji}_{\dot \g \dot\d)}\nonumber \\
&
+\ft1{3!2} \theta^{\a}_i \theta^{\b}_j \theta^{\g}_k \partial_{\a (\dot\a} \partial_{\b \dot\b}  \partial_{\g \dot \g} \bar{\chi}_{\dot \d)\ell m}\varepsilon ^{kji\ell m}+ \ft1{4!}\theta^{\a}_i \theta^{\b}_j \theta^{\g}_k \theta^{\d}_{\ell } \partial_{\a \dot\a} \partial_{\b \dot\b}  \partial_{\g \dot \g} \partial_{\d \dot \d}\phi _{m}\varepsilon^{ijk\ell m}\nonumber\\
&+\ft1{5!} \theta^{\a}_i \theta^{\b}_j \theta^{\g}_k \theta^{\d}_{\ell } \theta^{\eps}_{m} \partial_{\a \dot\a} \partial_{\b \dot\b}  \partial_{\g \dot \g} \partial_{\d \dot \d} \chi_{\eps} \varepsilon ^{ijk\ell m} \,,
  \label{chiralbarCN5}
\end{align}
where
\begin{equation}
  \phi _m=\phi _{m6}= \phi _{m678}\,,\qquad \chi _\alpha = \chi_{\alpha  678}\,,\qquad \bar\chi _{\dot \alpha ij}= \frac{1}{3!}\varepsilon _{ijk\ell m}\bar\chi _{\dot \alpha}^{k \ell m}\,,
 \label{redefinedN5}
\end{equation}
in terms of the fields that we had for ${\cal N}=6$ or ${\cal N}=8$. The set of fields is thus in this case
\begin{equation}
  \left\{{\cal N}=5\right\}= \left\{C_{\a\b\g\delta},\ \psi_{\a\b\g i},\  M_{\a\b ij},\ \chi_{\a}^{ ij},\ \phi^{i}, \ \bar{\chi}_{\dot{\a}}\right\}+ \cc\,.
\label{N5fields}
\end{equation}

Let us continue the truncation to ${\cal N}=4$. The superfield is now
\begin{align}
\bar{C}_{\dot\a\dot\b\dot\g\dot\d}(y, \theta)&= \bar C_{\dot\a\dot\b\dot\g\dot\d}(y) +\theta^{\a}_i  \partial_{\a (\dot{\a} } \bar{\psi}^i_{\dot \b \dot \g \dot \d)}
+\ft12\theta^{\a}_i \theta^{\b}_j \partial_{\a (\dot \a} \partial_{\b \dot\b} \bar{M}^{ji}_{\dot \g \dot\d)}
+\ft1{3!} \theta^{\a}_i \theta^{\b}_j \theta^{\g}_k \partial_{\a (\dot\a} \partial_{\b \dot\b}  \partial_{\g \dot \g} \bar{\chi}_{\dot \d)\ell }\varepsilon ^{kji\ell } \nonumber \\
&+ \ft1{4!}\theta^{\a}_i \theta^{\b}_j \theta^{\g}_k \theta^{\d}_{\ell } \partial_{\a \dot\a} \partial_{\b \dot\b}  \partial_{\g \dot \g} \partial_{\d \dot \d}\bar \phi \varepsilon^{ijk\ell }
 \,,
  \label{chiralbarCN4}
\end{align}
where in terms of ${\cal N}=5$ and ${\cal N}=8$ fields
\begin{equation}
  \bar{\chi}_{\dot \alpha i}= \bar \chi _{\dot \alpha i5}=-\frac{1}{3!}\varepsilon_{ijk\ell} \bar \chi ^{jk\ell}\,,\qquad \bar \phi =\phi _5=\phi _{5678}\,.
 \label{redefN4}
\end{equation}
The ${\cal N}=4$ fields are
\begin{equation}
  \left\{{\cal N}=4\right\}= \left\{C_{\a\b\g\delta},\ \psi_{\a\b\g i},\  M_{\a\b ij},\ \chi_{\a}^{ i},\ \phi\right\}+ \cc\,.
\label{N4fieldsapp}
\end{equation}
For ${\cal N}=3$ the superfield is
\begin{align}
\bar{C}_{\dot\a\dot\b\dot\g\dot\d}(y, \theta)&= \bar C_{\dot\a\dot\b\dot\g\dot\d}(y) +\theta^{\a}_i  \partial_{\a (\dot{\a} } \bar{\psi}^i_{\dot \b \dot \g \dot \d)}
+\ft12\theta^{\a}_i \theta^{\b}_j \partial_{\a (\dot \a} \partial_{\b \dot\b} \varepsilon ^{jik}\bar{M}_{\dot \g \dot\d)k}\nonumber\\
&+\ft1{3!} \theta^{\a}_i \theta^{\b}_j \theta^{\g}_k \partial_{\a (\dot\a} \partial_{\b \dot\b}  \partial_{\g \dot \g} \bar{\chi}_{\dot \d)\ell }\varepsilon ^{kji }
 \,,
  \label{chiralbarCN3}
\end{align}
where
\begin{equation}
  \bar{M}^{ij}_{\dot \alpha \dot \beta }= \varepsilon ^{ijk}\bar{M}_{\dot \alpha \dot \beta k}\,,\qquad \bar{\chi}_{\dot \alpha  }=\bar{\chi}_{\dot \alpha 4 }=\frac{1}{3!}\varepsilon _{ijk}\bar \chi _{\dot \alpha }^{ijk}\,.
 \label{redefN3}
\end{equation}
The ${\cal N}=3$ fields are
\begin{equation}
  \left\{{\cal N}=3\right\}= \left\{C_{\a\b\g\delta},\ \psi_{\a\b\g i},\  M_{\a\b}^ i,\ \chi_{\a}\right\}+ \cc\,.
\label{N3fields}
\end{equation}
The ${\cal N}=2$ superfield is
\begin{align}
\bar{C}_{\dot\a\dot\b\dot\g\dot\d}(y, \theta)&= \bar C_{\dot\a\dot\b\dot\g\dot\d}(y) +\theta^{\a}_i  \partial_{\a (\dot{\a} } \bar{\psi}^i_{\dot \b \dot \g \dot \d)}
+\ft12\theta^{\a}_i \theta^{\b}_j \partial_{\a (\dot \a} \partial_{\b \dot\b} \varepsilon ^{ji}\bar{M}_{\dot \g \dot\d)}
 \,,
  \label{chiralbarCN2}
\end{align}
where
\begin{equation}
 \bar{M}_{\dot \alpha  \dot\beta }= \bar{M}_{\dot \alpha  \dot\beta 3} \,.
 \label{redefN2}
\end{equation}
The ${\cal N}=2$ fields are
\begin{equation}
  \left\{{\cal N}=2\right\}= \left\{C_{\a\b\g\delta},\ \psi_{\a\b\g i},\  M_{\a\b}\right\}+ \cc\,.
\label{N2fields}
\end{equation}
The ${\cal N}=1$ superfield is
\begin{equation}
\bar{C}_{\dot\a\dot\b\dot\g\dot\d}(y, \theta)= \bar C_{\dot\a\dot\b\dot\g\dot\d}(y) +\theta^{\a}  \partial_{\a (\dot{\a} } \bar{\psi}_{\dot \b \dot \g \dot \d)}
 \,,
  \label{chiralbarCN1}
\end{equation}
in terms of the field content
\begin{equation}
  \left\{{\cal N}=1\right\}= \left\{C_{\a\b\g\delta},\ \psi_{\a\b\g }\right\}+ \cc\,.
\label{N1fields}
\end{equation}
For ${\cal N}=0$ there is just the field $\bar{C}_{\dot\a\dot\b\dot\g\dot\d}(x)$ and its complex conjugate. We do not know how to continue for ${\cal N}<0$, and therefore we stop here.

\subsection{Chiral superfields \texorpdfstring{$\Phi_{\cN}$}{} for \texorpdfstring{$7\geq\cN\geq1$}{N=7 to 1}.}

The highest component of a chiral superfield is the lowest component of an anti-chiral multiplet. See e.g. the chiral multiplets in ${\cal N}=1$, which have components $\{Z,\,P_L\chi,\,F\}$. The field $F$ transforms only with $P_R\epsilon $ and defines therefore an anti-chiral multiplet with components $(F,\,\slashed{D}P_L\chi ,\, \Box Z)$.

We will prove below that we can say the same for the chiral superfields starting with $\bar C_{\dot \alpha \dot \beta \dot \gamma \dot \delta }$ mentioned above for ${\cal N}=8,\ldots ,0$. This highest component is always a singlet under $\SU({\cal N})$ and defines an anti-chiral multiplet.  The complex conjugate of the latter is then again a chiral superfield.
This leads to the list of (anti)chiral superfields in table~\ref{tbl:chiralN80}.

For ${\cal N}=8$ the statement is simple.
The statement of anti-chirality of $C_{\alpha \beta \gamma \delta }$ is just the complex conjugate of the statement of chirality  of $\bar C_{\dot \alpha \dot \beta \dot \gamma \dot \delta }$.

For ${\cal N}= 7,6,5,4$ the anti-chirality of the fields $\psi _{\alpha \beta \gamma 8}$, $M_{\alpha \beta 78}$, $\chi _{\alpha 678}$ and $\phi _{5678}$  is proven immediately from  (\ref{cn1}). E.g. for ${\cal N}=7$, the anti-chirality means that the superfield vanishes under $D^i_\alpha $ for $i=1,\ldots ,7$. For these values of $i$ and $j=8$ in the last equation of (\ref{cn1}), the result is immediate. The other 3 equations of  (\ref{cn1}) imply the same result for ${\cal N}=6,5,4$.

For ${\cal N}=3$ we prove the chirality of $\chi _{\alpha 123}$ from the first line of (\ref{barDchi}). Indeed, in the right-hand side comes then $\phi _{i123}$ with $i=1,2,3$, and this thus vanishes.
In the same way, the next two lines of (\ref{barDchi}) prove the chirality of $M_{\alpha \beta 12}$ for ${\cal N}=2$, and of $\psi _{\alpha \beta \gamma 1}$ for ${\cal N}=1$.

We can give explicit components of these superfields by using the complex conjugates of (\ref{barDchi}).
%The chiral superfield $\bar{M}^{78}_{\dot{\a}\dot{\b}}$ for ${\cal N}=6$ is defined by
%\be
%\bar{M}^{78}_{\dot{\a}\dot{\b}}=\bar{D}_{i\dot{\a}}\bar{D}_{j\dot{\b}}\phi^{ij}.
%\ee
This gives
\begin{eqnarray}
\bar{M}^{78}_{\dot{\a}\dot{\b}}(y,\theta)&=&\bar{M}^{78}_{\dot{\a}\dot{\b}}+\theta^\a_i\partial_{\a(\dot{\a}}\bar{\chi}^{i78}_{\dot{\b})}
+\ft12\theta^\a_i\theta^\b_j\partial_{\a\dot{\a}}\partial_{\b\dot{\b}}\phi^{ji}
-\ft1{3!3!}\theta^\a_i\theta^\b_j\theta^\g_k\partial_{\a\dot{\a}}\partial_{\b\dot\b}\varepsilon ^{ijk\ell m n}\chi_{\g \ell m n}\nonumber\\
&&-\ft1{4!2}\theta^\a_i\theta^\b_j\theta^\g_k\theta^\delta_\ell\varepsilon^{ijk\ell mn}\partial_{\a\dot{\a}}\partial_{\b\dot\b}M_{\g\delta mn}
-\ft1{5!}\theta^\a_i\theta^\b_j\theta^\g_k\theta^\delta_\ell\theta^\eta_m\varepsilon^{ijk\ell mn}\partial_{\a\dot{\a}}\partial_{\b\dot\b}\psi_{\g\delta\eta n}\nonumber\\
&&-\ft1{6!}\theta^\a_i\theta^\b_j\theta^\g_k\theta^\delta_\ell\theta^\eta_m\theta^\sigma_n\varepsilon^{ijk\ell mn}\partial_{\a\dot{\a}}\partial_{\b\dot\b} C_{\g\delta\eta\sigma}.
 \label{M78D6}
\end{eqnarray}

For ${\cal N}=5$, in terms of $\phi$, the conditions \eqref{cn1}, \eqref{cn2} are
\begin{eqnarray}
D_\a^i\phi^j&=&\frac{1}{3!}\varepsilon^{jik\ell m}\chi_{\a k\ell m},\nonumber\\
\bar{D}_{i\dot{\a}}\phi^j&=&\delta^i_j\bar{\chi}_{\dot{\a}}^{678}.
 \label{N5Bianchi}
\end{eqnarray}
%We have defined $\bar{\chi}_{\dot{\a}}=\frac{1}{5!}\varepsilon ^{ijk\ell m}\bar{\chi}_{\dot{\a}ijk\ell m}$ which is a chiral superfield.
With the chiral bases, we can expand $\bar{\chi}_{\dot{\a}}^{678}$ as
\begin{eqnarray}
\bar{\chi}_{\dot{\a}}^{678}(y,\theta)&=&\bar{\chi}^{678}_{\dot{\a}}+\theta_i^\a\partial_{\a\dot{\a}}\phi^i+\ft12\theta_i^{\a}\theta_j^\b\partial_{\a\dot{\a}}\chi_\b {}^{ij} +\ft1{3!2}\theta_i^\a\theta_j^\b\theta_k^\g \varepsilon ^{ijk\ell m} \partial_{\a\dot{\a}}M_{\b\g lm}\nonumber \\
&&+\ft1{4!}\theta_i^\a\theta_j^\b\theta_k^\g\theta_\ell^\delta\varepsilon ^{ijk\ell m} \partial_{\a\dot{\a}}\psi_{\b\g\delta m}+\ft1{5!}\theta_i^\a\theta_j^\b\theta_k^\g\theta_\ell^\delta\theta_m^\eta\varepsilon ^{ijk\ell m} \partial_{\a\dot{\a}}C_{\b\g\delta\eta}.
 \label{barchiD5}
\end{eqnarray}

In $\cN=4$ we have
\bea
W(y, \theta) &=& \phi(y)- \theta^{\a}_i  \chi_\alpha ^i +\ft1{2\cdot 2}\theta^{\a}_i \theta^{\b}_j {M}_{k\ell \a \b} \varepsilon^{ijk\ell } +\ft1{3!} \theta^{\a}_i \theta^{\b}_j \theta^{\g}_k {\psi}_{\a \b \g\ell }\varepsilon ^{ijk\ell} \cr
&&+ \ft1{4!}\theta^{\a}_i \theta^{\b}_j \theta^{\g}_k \theta^{\d}_{\ell } C_{\a\b\g\delta} \varepsilon^{ijk\ell }\,.
\eea

In $\cN=3$, we have
\begin{eqnarray}
W_\a (y, \theta)= \chi_\a(y)+\theta^\b_iM_{\a\b}^i(y)+\frac{1}{2}\theta^\b_i\theta^\g_j\varepsilon^{ijk} \psi_{\a\b\g k}(y) +\frac{1}{3!}\theta^\b_i\theta^\g_j\theta^\d_k\varepsilon^{ijk} C_{\a\b\g\d}(y).
\end{eqnarray}

In $\cN=2$,
\begin{eqnarray}
W_{\a\b} (y, \theta)=M_{\a\b}(y)+\theta^\g_i\varepsilon^{ij} \psi_{\a\b\g j}(y) +\frac{1}{2}\theta^\g_i\theta^\d_j\varepsilon^{ij} C_{\a\b\g\d}(y).
\end{eqnarray}

In $\cN=1$,
\begin{eqnarray}
W_{\a\b\g} (y, \theta)= \psi_{\a\b\g}(y) +\theta^\d C_{\a\b\g\d}(y).
\end{eqnarray}

\subsection{Completeness of the list of chiral superfields  }
It is important to explain the reason why our list of linearized chiral superfields in table~\ref{tbl:chiralN80} is complete.
We are trying to detect all possible $\U(1)$ anomalous 4-point amplitudes, we construct them using chiral linearized superfields. If our list would be incomplete, we would not be able to provide a complete list of candidates into anomalous amplitudes.

A straightforward argument about completeness is the following. One can look at all $\cN\geq 0$ pure supergravities which have known actions and known local supersymmetry rules. In general, for all component fields $\Phi(x)$ the supersymmetry transformation of the action has the form
\be
\delta_s \Phi(x)= \epsilon_i^\alpha  \Psi^i _\alpha(x) + \bar \epsilon^{i \dot \alpha} \Theta_{i \dot \alpha}(x).
\label{susy}\ee
Here we have suppressed the possible indices of the field $\Phi$ and a dependence on them in $\Psi$ and $\Theta$. In extended supergravities with $\cN \geq 4$ both $\Psi$ and $\Theta$ terms are present. This means that at the non-linear level for all component fields of the theory $\Psi\neq 0$ and $\Theta\neq 0$, there no chiral superfields. Starting with $\cN=3$ there is a non-vanishing superspace torsion which breaks the integrability of the chirality constraint,
\be
D_\alpha ^i \Phi (x, \theta, \bar \theta) =0.
\ee
This condition is inconsistent with the algebra of covariant derivatives
\be
\{ D_\alpha ^i , D_\beta ^j \} \Phi (x, \theta, \bar \theta) = T_{\alpha \, \beta}^{\dot \alpha\, ijk } \bar D_{k \dot \alpha} \Phi (x, \theta, \bar \theta)+ \cdots.
\label{algebra}\ee
The chiral superfield must also be an anti-chiral one, and therefore constant. The torsion superfield $T_{\alpha \, \beta}^{\dot \alpha\, ijk }= \epsilon_{\alpha \beta} \bar \chi^{\dot \alpha\, ijk } (x, \theta, \bar \theta) $ has  a spin 1/2 field $\bar \chi^{\dot \alpha\, ijk } (x)$ as its first component.

However, once we are interested in the linearized superfields, the r.h.s. of \rf{algebra} can be neglected, being at least quadratic in component fields. This means in terms of linearized supersymmetry transformations that one of the entries in the r.h.s.  of \rf{susy} vanishes and we find linearized chiral superfields whose first component is $\Phi(x)$ if
\be
\Big (\Theta_{i \dot \alpha}(x)\Big )_{\rm lin}=0,
\ee
and anti-chiral superfields whose first component has
\be
\Big ( \Psi^i _\alpha(x)\Big )_{\rm lin}=0.
\ee
The list of all available linearized chiral superfields shown in table~\ref{tbl:chiralN80} has been established by a direct inspection of linearized  supersymmetry transformations in $\cN\geq 0$ supergravities.

%%%%%%%%%%%%%%%%%%%%%%%%%%%%%%%%%%%%%%%%%%%

\bibliographystyle{JHEP}
\bibliography{refs}

\end{document}